%% file: Phasetype_Influence.tex
\documentclass[preprint,12pt,authoryear]{elsarticle}

\usepackage{float}
\usepackage{amssymb}
\usepackage{amsmath}
\usepackage{graphicx}
\usepackage{tikz}
\usepackage{subcaption}
\usepackage{xcolor}
\usepackage{makecell}
\usepackage[hidelinks]{hyperref}
\usepackage{tikz}
\usetikzlibrary{shapes.geometric, arrows}

\usepackage{siunitx}

\usepackage[linesnumbered,ruled,vlined]{algorithm2e}

\include{rwth-colors}
\usetikzlibrary{arrows.meta}

\begin{document}
\begin{frontmatter}

\title{Utilizing phase-type distributions for queueing-based railway junction performance determination}

\author[inst1]{Tamme Emunds}
\ead{emunds@via.rwth-aachen.de}

\affiliation[inst1]{organization={Institute of Transport Science,\\ RWTH Aachen},
            addressline={{Mies-van-der-Rohe Straße 1}}, 
            city={Aachen},
            postcode={52074},
            country={Germany}}

\author[inst1]{Nils Nießen}

\begin{abstract}
To ensure the effective and objective development of transportation networks, it is crucial to identify performance limitations across various subsystems. A timetable-independent assessment of infrastructure capacity at railway junctions is a fundamental aspect of long-term rail network planning. 
While recent research introduced queueing-based methods to quantify route-based railway junction performance, modelling arrival and service processes has been limited to exponential distributions. This work utilizes Phase-Type Distributions to propose an extension to a previously described Continuous-Time Markov Chain model. In a comparison between assumed distribution combinations, the effect of a more detailed stochastic process modelling is described.
Furthermore, an analysis of the differences to a simulation method is conducted for an exemplary railway junction.
The introduced method enables infrastructure managers to accurately model stochastic processes for performance determination in the early stages of the strategic planning phase.

\end{abstract}


\end{frontmatter}


\section{Introduction}
\label{sec:intro}


Due to the increasing demand for environmentally friendly modes of transportation, railway infrastructure managers worldwide need to develop their infrastructure further.
Performance analysis of individual facilities serves as a valuable method to evaluate existing infrastructure and compare expansion and new construction scenarios.
Some well established methods are particularly suited for analysing current infrastructure and timetables, for example those designed to determine capacity utilisation \citep{UIC.2004}.
On the other hand, efficient resource management requires a focus on long-term planning.
For this purpose, timetable-independent methods can be particularly useful.

While such methods for railway lines are already well established in their practical implementation, methods for railway junctions or nodes are still the subject of ongoing research.
The currently utilized systems in practice employ single-channel approximations for multichannel service systems, such as railway route nodes and junctions.
However, those systems with parallel utilisable routes are particularly critical points that need to accommodate the traffic from multiple railway lines.

In a preceding work \citep{emundsEvaluatingRailwayJunction2024}, a multichannel method was introduced to analyze the performance of railway junctions and estimate the necessity for an overpass structure to mitigate route dependencies.
In that study, the arrival and service processes were modeled using only exponential distributions, necessitating the use of approximation formulas to accommodate other probability distributions.

In contrast, this work introduces a novel approach, modelling general independent arrival and service processes with phase-type distributions in a Continuous-Time Markov Chain for multi-channel railway systems.
This model can be used to obtain more accurate approximations of the queue-lengths of the different routes through the infrastructure, thereby enabling timetable-independent quantifications of the occupancy of each route, which allows for a detailed bottleneck analysis of the infrastructure.

The contribution of this work is manifold, including:
\begin{itemize}
    \item A novel Continuous-Time Markov Chain model that utilizes phase-type distributions for modeling arrival and service processes in a railway junction.
    \item An algorithm designed to efficiently compute timetable capacity by comparing queue-length estimations to their respective thresholds.
    \item A comprehensive study analyzing the influence and approximation quality of different queuing models for varying distributions of traffic types across routes.
    \item A case study demonstrating the usability of route-based analysis to evaluate the performance of a railway junction infrastructure that accommodates both freight and passenger traffic.
\end{itemize}

The remainder of this work is structured as follows. 
Section \ref{sec:related_work} presents the current state of research. 
In Section \ref{sec:problem_descr}, a formal description of the junction capacity determination problem is provided. 
Subsequently, the novel model and the employed approximation formulas are introduced in Sections \ref{sec:approx_formel} to \ref{subsec: Queue-Length Analysis}, along with an explanation of the algorithm for efficient capacity determination in Section \ref{subsec:cap_deter_alg}. 
Section \ref{sec:validation} offers a validation of the proposed method and a comparison with other models under different traffic distribution scenarios. 
Finally, an example railway junction combining a freight and a passenger traffic line is analyzed in Section \ref{sec:case_study}.

\section{Related Work}
\label{sec:related_work}

This section provides a summarized overview of the state-of-the-art for railway capacity analysis, focusing on key methodologies and their applications.

Railway capacity is defined in various ways depending on the planning stage and requirements \citep{Jensen.2020, emundsEvaluatingRailwayJunction2024}.
The \textit{theoretical capacity} refers to the maximum number of trains that can be scheduled without conflicts, considering driving dynamics and control systems.
The \textit{timetable capacity} or \textit{maximal capacity} includes additional factors such as train-mix and schedule quality.
Finally, the \textit{operational capacity} or \textit{practical capacity} accounts for disturbances and delays, ensuring acceptable operational quality.
Research also focusses on calculating the \textit{capacity utilization}, which measures how much of the available capacity is used in a given timetable.

Various methodologies have been employed for railway performance analysis and vary in their dependency on timetables; some are timetable-dependent while others are not, making them useful for early infrastructure planning stages.

These methodologies differ based on the analyzed infrastructure (lines, junctions, stations, networks), infrastructure decomposition, and solution methods like mixed integer programming (MIP) or matrix calculations.

The \textit{UIC Code 406} (\citet{UIC.2004,uicCode406Capacity2013}) is widely used internationally for assessing railway line and station capacities by determining capacity utilization through compression methods requiring a timetable or randomly generated sequences of train types to overcome timetable dependencies (\citet {Goverde.2007, Goverde.2013,Abril.2008, Besinovic.2018, Jensen.2020}).

\textit{Optimisation methods} estimate theoretical capacity through linear mixed-integer programming problems for lines, stations, and networks.
They solve railway timetabling problems to build optimal timetables based on objective functions (\citet{Zwaneveld.1996, Burdett.2006, Burdett.2016, Harrod.2009, Lusby.2011, Cacchiani.2012, Cacchiani.2016, Yaghini.2014}).
Some optimisation approaches incorporate rolling stock information or model delay propagation effects using max-plus algebras (\citet {Kort.2003,Mussone.2013,Liao.2021}).

\textit{Simulations} provide detailed insights into operational parameters by simulating train operations based on given timetables or random variables representing arrival and service processes (\citet{Zieger.2018, Dacierno.2019}).

\textit{Analytical methods} based on queueing theory efficiently analyze timetable or operational capacities during early planning stages by considering inter-arrival and service time distributions.
Methods to determine the performance of railway lines (\citet{Schwanhauer.1974, Schwanhauer.1982, Wendler.2007, Weik.2017}),  railway junctions (\citet{Schwanhauer.1978, Niessen.2008, Niessen.2013, Schmitz.2017, Weik.2020PhD, emundsEvaluatingRailwayJunction2024}) and track groups (\citet{GerhartPotthoff.1970, fischer1990bedienungsprozesse}) have been introduced.
Table \ref{table:lit_ana} compares other queueing-based methods for evaluating railway junction performance with the model introduced here.

\begin{table}[!htp]
\centering
\caption{Analytical junction capacity determination methodology}
\label{table:lit_ana}
\resizebox{\textwidth}{!}{
\begin{tabular}{c||c|c|c|c|c|c}
\thead{Literature} & \thead{multi-channel \\ model} &\thead{timetable \\ capacity}& \thead{route-based \\ decomposition} & \thead {route-based \\ quality \\ assessment} & \thead{explicit \\ non exponential \\ process modelling} & \thead{solution \\ method} \\ \hline \hline
\citep{Schwanhauer.1978}&    & \checkmark    &     (\checkmark)                        &    &         & \makecell{closed-form \\ formula} \\ \hline
\citep{Niessen.2008, Niessen.2013}& \checkmark   & \checkmark    &                            & (\checkmark)    &         & \makecell{iterative \\ formula} \\ \hline
\citep{Schmitz.2017}              & \checkmark   & \checkmark    &                            &                 & \checkmark  & \makecell{matrix-vector \\ equations} \\ \hline
\citep{Weik.2020PhD}   &   & \checkmark    &  (\checkmark)    &           &   \checkmark & \makecell{matrix-vector \\ equations} \\ \hline 
\citep{emundsEvaluatingRailwayJunction2024}   & \checkmark   & \checkmark    &  \checkmark      &    \checkmark         &             & \makecell{probabilistic \\ model-checking} \\ \hline\hline
\makecell{introduced \\ here}      & \checkmark  & \checkmark    &  \checkmark                & \checkmark      &  \checkmark  & \makecell{probabilistic \\ model-checking} \\ \hline  \hline
\multicolumn{7}{p{1.4\textwidth}}{Remarks: Please note that a statement in bracelets means that this feature is partially supported.}
\end{tabular}
}
\end{table}

For example, in \citet{Niessen.2008, Niessen.2013}, interlocking nodes are analysed for their timetable and operational capacity using loss probabilities in service systems. To this end, exponentially distributed inter-arrival and service times are employed and results are scaled to approximate non-exponential service times, utilizing the variation coefficient.

In \citet{Schmitz.2017}, a railway junction is divided into two service stations and is modelled as a Markov Chain. Notably, the exponential scaling of the state space due to the storage of the request type, as well as the assumption of an existing subdivision into independent service stations, are significant properties of this approach.
This method allows a railway junction to be considered as a multi-channel service system and analysed using phase-type distributed arrival and service processes.

To analyse the significance of switch connections in station yards, \citet{Weik.2020PhD} examines interlocking nodes by approximating them as a series of interconnected service stations in a single-channel system, following the methodology of \citet{Schwanhauer.1978}. In this model, the arrival, service, and an additional repair processes are represented using phase-type distributions.

In contrast, this paper is based on the model proposed by \citet{emundsEvaluatingRailwayJunction2024}.
This model enables the performance evaluation of various junction infrastructures by discretizing the utilized infrastructure based on operational route paths and modeling the arrival and service of requests on this infrastructure using a Markov chain.
Initially, only exponentially distributed inter-arrival and service processes were considered using the Markov chain model, which had to be converted into processes with variation coefficients not equal to 1 using approximation formulas for multi-channel service systems according to \citet{fischer1990bedienungsprozesse}.

In this work, models are derived that enable the computation of phase-type distributions. In particular, several influencing factors on the accuracy of the approximation are presented and analyzed.

\section{Problem Description}
\label{sec:problem_descr}

This work introduces a method to evaluating the performance of a double-track railway junction.
The performance is assessed using the metric known as \textit{timetable capacity}, which gauges the infrastructure's ability to accommodate train schedules, see also \citet{Wendler.2007, emundsEvaluatingRailwayJunction2024}.
Specifically, it pertains to the timetabling process, wherein the infrastructure operator allocates service slots to various requests submitted by transport operators.
With this approach, random distributed requests can be modelled, encompassing all potential timetables that could result in such a procedure.
Conflicts may arise due to the inherent randomness of the requests, which might necessitate imposing waiting times for some of these requests.
Furthermore, it may be necessary to decline certain requests if no conflict-free allocation is possible.
The performance of the infrastructure can be assessed by the estimated number of requests that are pending, referred to as the \textit{estimated queue length} $L$.
Although the method we propose is designed to model the allocation of scheduling requests, for clarity, we often refer to these requests as \textit{trains}.

A \textit{railway junction} describes the infrastructure that connects railway lines to more than two different destinations.
It usually consists of entry- and exit-signals from and to every direction, as well as one or multiple switches to set the path for the different routes between origins and destinations.
In comparison with a \textit{railway station}, trains are not allowed to have scheduled stops in a railway junction, they only stop if the signalling system indicates occupied infrastructure in their planned route.

\begin{figure}[ht]
    \centering
    \includegraphics[width=\textwidth]{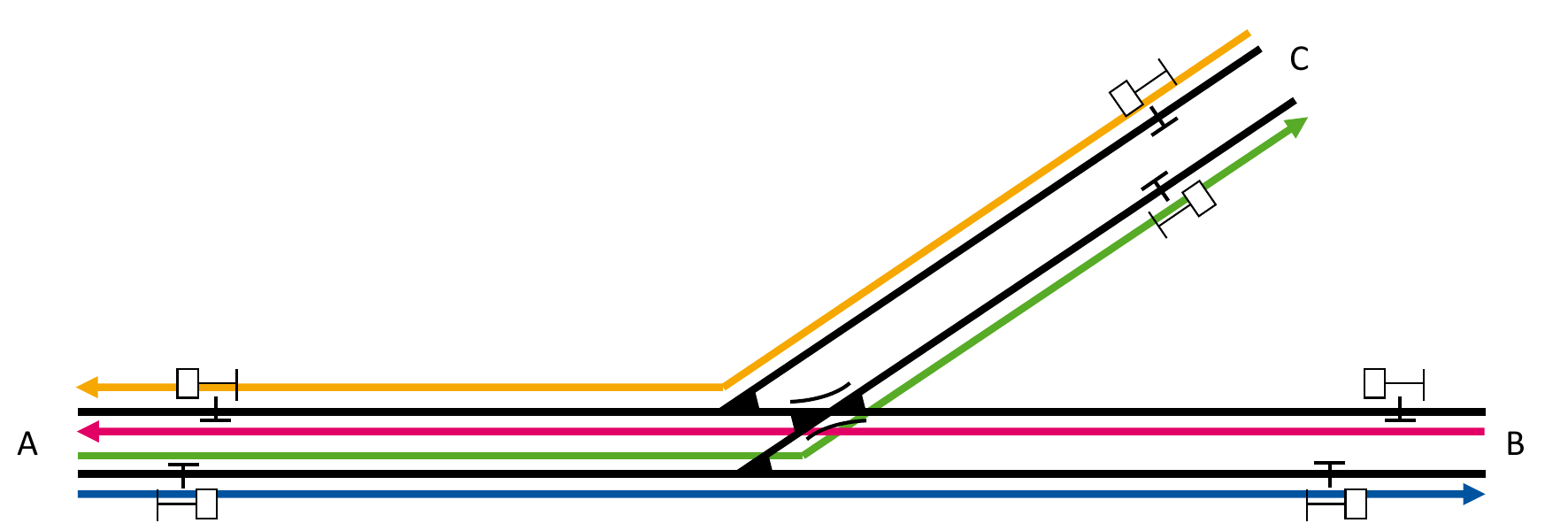}
    \caption{Examplary infrastructure of a railway junction}
    \label{fig:junction}
\end{figure}

However, in the remainder of this work, a simple example of a double-track railway junction with four different routes will be considered.
These routes can be distinguished regarding their direction, \textcolor{rwth}{A-B ($r_1$)} and \textcolor{grun}{A-C ($r_2$)} start in A towards B or C, while \textcolor{magenta}{B-A ($r_3$)} and \textcolor{orange}{C-A ($r_4$)} end in A, coming from B or C.
An exemplary infrastructure is depicted in Figure \ref{fig:junction}.

Since no use of turnout tracks is required for the two routes between A and B, we refer to the line between A and B as the \textit{main line} and the line between A and C as the \textit{branch line}.
The proportion of the main line in the total traffic volume is then given by 
\begin{equation}
    p_{main} = \frac{n_{main}}{n_{total}},
\end{equation}
where it depends on the number of trains on the main line ($n_{main}$) and the total number of trains  $n_{total}= n_{main} + n_{branch}$ on main and branch line ($n_{branch}$).

We express the railway junction as a tuple $J = \left(R, C\right)$ of a set $R$ of $k$ routes and a \textit{conflict matrix} $C \in \{0,1\}^{k \times k}$, describing whether two routes $r, r^{\prime} \in R$ can be used at the same time ($C_{r, r^{\prime}} = 0$), or are \textit{conflicting} ($C_{r, r^{\prime}} = 1$).

In addition to the infrastructure, performance can depend on the used train types $T$.
We therefore formulate an occupation request type $o = (r, t) \in O \subset R \times T$ as a combination of a route $r \in R$ and a train $t \in T$.

For some examples, it might be suitable to give the distribution of  a given total number $n_{\text{total}}$ of requests to the number of requests of same type $o=(r, t)$.
This can be formalized by a function
\begin{equation}
    \theta: \mathbb{R} \rightarrow \mathbb{R}^{|O|}: n_{\text{total}} \mapsto \theta\left(n_{\text{total}}\right),
\end{equation} 
yielding the number of requests $n_{r, t} = \theta_{r, t}\left(n_{\text{total}}\right)$ for every occupation request type $o= (r, t)$.

Determining the timetable capacity of a railway junction involves the definition and calculation of multiple parameters, some of which can already serve as an indicator about the quality of the planned junction. 
Table \ref{tab:notations} gives a list of the used notations.

\begin{table}[htp]
    \centering
    \caption{List of notations.}
    \resizebox*{!}{0.95 \textheight}{
    \begin{tabular}{|c|c|}\hline
        $R$ & Set of routes  \\ 
        $r_i$ & Route $i$ \\ 
        $C \in \{0,1\}^{k \times k}$ & Conflict matrix \\ 
        $J=(R, C)$ & Junction infrastructure\\  
        $t_U$ & Time horizon \\
        $T$ & Set of trains \\
        $O$ & Set of occupation request types\\
        $\theta$ & Request distribution \\\hline
        $\lambda$ & Arrival rate \\ 
        $\mu$ & Service rate \\ 
        $\rho$ & Occupancy rate \\ 
        $v_A$ & Variation coefficient arrival process \\ 
        $v_S$ & Variation coefficient service process \\
        $S$ & Set of states \\ 
        $q_i$ & Number of requests in queue of route $r_i$ \\ 
        $s_i$ & Status of Service of route $r_i$ \\ 
        $p_{A,i}$ & Phase of arrival process on route $r_i$ \\
        $p_{S,i}$ & Phase of service process on route $r_i$ \\
        $M$ & Maximum rate \\ 
        $MC = \left(S, T\right)$ & Continuous-Time Markov Chain \\ 
        $T$ & Set of transitions \\ 
        $t=(u, v)$ & Transition from state $u$ to state $v$\\
        $L_r$ & Expected queue length of route $r$ \\ 
        $m$ & number of waiting slots \\ \hline
        $L_{limit}$ & Threshold value for sufficient quality \\ 
        $p_{\text{pt}}$ & Share of passenger trains \\
        $p_{\text{main}}$ & Share of main line traffic\\
        $n_{\text{total}}$ & Total number of trains \\
        $n_{\text{max}}$ & Timetable capacity \\ \hline
        $p_{\text{suburban}}$ & Share of suburban trains \\
        $p_{\text{regional freight}}$ & Share of regional freight trains \\
        $n_{\left(r, t\right)}$ & Train numbers on route $r$ and for traffic type $t$\\
        $h_{i,j}$ & Minimum headway time of the sequence train $j$ after train $i$ \\
        $b$ & Service time \\
        $n_r$ & Number of trains on route $r$ \\  \hline
        \end{tabular}%
        }
    \label{tab:notations}
\end{table}


Performance analysis of railway infrastructure is typically conducted with respect to a fixed \textit{time horizon} $t_U$, which defines the duration of the investigation period in minutes. The \textit{arrival rate} $\lambda_r = \frac{n_r}{t_U}$ represents the average number of trains per route that request service on route $r$ per minute.

The performance of railway infrastructure is significantly influenced by the service times planned for train usage. For this purpose, the \textit{minimum headway times} $h_{(r_i, t_i), (r_j, t_j)}$ describe the minimum time interval required for the route-train combination $(r_j, t_j)$ to initiate service after the service of the preceding route-train combination $(r_i, t_i)$ has commenced.
These headway times are dependent not only on the infrastructure and the specific routes that the trains $t_i$ and $t_j$ are scheduled to traverse, but also on certain rolling stock parameters, such as acceleration and braking behavior, as well as train length. 

For comprehensive details on obtaining blocking and minimum headway times, readers are referred to \citet{hansen2014railway}. In this paper, it is assumed that the minimum headway times are provided; typically, these would have been calculated using a microscopic tool prior to conducting a detailed infrastructure analysis.

Let $n_{r,t}$ denote the number of trains on route $r$ of train type $t$.
For a given pair $r, r^{\prime}$ of conflicting routes ($C_{r, r^{\prime}} = 1$), the minimum headway time $h_{(r, t), (r^{\prime}, t^{\prime})}$ of all possible sequences $(r, t), (r^{\prime}, t^{\prime})$ of train-route combinations for this route pair $r, r^{\prime}$, can be weighted with the total number of possible sequences for $r$ and $r^{\prime}$,
$n_{r, r^{\prime}} = \sum_{t} \sum_{t^{\prime}} n_{r,t} \cdot n_{r^{\prime},t^{\prime}}$, to obtain the average minimum headway time 
\begin{equation}
    h_{r, r^{\prime}} = \sum_{t} \sum_{t^{\prime}} n_{r,t} \cdot n_{r^{\prime},t^{\prime}} \cdot h_{(r, t), (r^{\prime}, t^{\prime})} \cdot \frac{1}{n_{r,r^{\prime}}}
\end{equation}

between all train pairs $t, t^{\prime}$ on route $r$ and $r^{\prime}$ respectively.
By further weighting the average minimum headway times $h_{r, r^{\prime}}$ for each conflicting route pair with the total number of trains $n_{r^{\prime}}$ per conflicting route $r^{\prime}$, the average \textit{service time} for a route $r$ can be calculated as:
\begin{equation}
    b_{r} = \sum_{\substack{r^{\prime} \\ C_{r, r^{\prime}} = 1}} \frac{n_{r^{\prime}}}{n_{r, \text{conflict}}} \cdot h_{r, r^{\prime}},
\end{equation}
where $\frac{n_{r^{\prime}}}{n_{r, \text{conflict}}}$ describes the probability of the pair $(r, r^{\prime})$, i.e. a sequence describing any train on the route $r^{\prime} \in R$ following any train on the conflicting route $r$. 
Here,  the total number of conflicting trains for route $r$ is given by
\begin{equation}
    n_{r, \text{conflict}} = \sum_{\substack{r^{\prime} \\ C_{r, r^{\prime}} = 1}} n_{r^{\prime}}.
\end{equation}

Using this service time $b_r$ for each route, the \textit{service rate} $\mu_r = \frac{1}{b_r}$ describes the average number of trains that can be serviced on route $r$ per minute.

The \textit{occupation ratio} $\rho_r$ of a route can be determined by comparing the arrival- and service rates
\begin{equation}
    \rho_r = \frac{\lambda_r}{\mu_r}.
\end{equation}
Furthermore, to determine the timetable capacity of a railway junction, the expected queue-length $L_{r}$ can be calculated for every route $r \in R$ utilizing the arrival rate $\lambda_r$ and service rate $\mu_r$.
In this work, the notation $EL_r$ is sometimes chosen, when explicitly referring to approximations of this expected queue-length. 


To enforce a sufficient timetable quality, a limit to the expected queue length $L_{\text{limit}, r}$ can be set by the infrastructure manager, see Section \ref{subsec:threshold} for a detailed description.
This limit depends on the share of passenger trains and can therefore vary between every route, depending on its operating program.
However, the expected queue-length of every route can be compared to the route-specific 
threshold value to determine the quality factor
\begin{equation}
    qf_r = \frac{L_r}{L_{\text{limit},r}}
\end{equation}
of this specific route.
Therefore, the bottleneck in the analysed railway junction infrastructure layout can be identified by comparing the quality factors of the different routes.

The timetable capacity of a given railway junction infrastructure $J$ can be formulated as the maximal number of train requests $n_{\text{max}}$ that can be scheduled, under consideration of a fixed distribution $\theta$ to routes, while respecting given minimum headway times $h_{(r_i, t_i), (r_j, t_j)}$ and not exceeding a set limit $L_{\text{limit}, r}$ of the expected queue-length per route.


A method to calculate the route-based queue length has been introduced in \cite{emundsEvaluatingRailwayJunction2024}. It corresponds to modelling the railway junction as a Continuous-Time Markov Chain (CTMC) and calculating the state probabilities utilizing probabilistic Model-Checking.


Since this model utilizes Markov Chains to model the arrival and service processes, it assumes exponentially distributed inter-arrival and service times.
Therefore, the arrival and service processes have been supposed to be completely random, expressed by their coefficient of variation of $v_A = v_B = 1$. In correspondence with the Kendall notation (see \cite{Kendall.1953}), we call this fully exponential setting (M/M).

However, the coefficient of variation can vary between different settings and general planning rules specify the coefficient of variation for the arrival process to $v_A = 0.8$ and for the service process to $v_B = 0.3$.
Hence, more general distributions need to be analysed in a queueing system for railway infrastructure.
We call this general independent setting (GI/GI), also following the Kendall notation.

The next section proposes a novel Continuos-Time Markov Chain model using phase-type distributions and further explains quality thresholds.

\section{Methods}

In this section, the methods for calculating the timetable capacity of a railway junction are presented. 
First, the threshold values that ensure sufficient infrastructure quality are defined in Section \ref{subsec:threshold}. 
Next, the approximation methods used to adapt exponential distributions to  general independent settings are discussed in Section \ref{sec:approx_formel}.
The novel model, which incorporates phase-type distributions for the arrival and service processes, is introduced in Section \ref{subsec:phase_type}. 
Finally, the algorithm designed to efficiently compute timetable capacity using queue-length estimations (Section \ref{subsec: Queue-Length Analysis}) is described in Section \ref{subsec:cap_deter_alg}.

\subsection{Threshold Values}
\label{subsec:threshold}

The calculated queue-length $EL_r$ for a route can be compared with limit values for the queue-length $L_{limit, r}$ to assess the quality of the provided transport service.
For long-term infrastructure planning, the largest german infrastructure manager, \citet{DBNetzAG.2009}, utilizes the threshold
\begin{equation}
\label{wsl_limit_sh}
  L_{\text{limit, r}}  = 0.479 \cdot \mathrm{exp}(-1.3 \cdot p_{\text{pt}, r})
\end{equation}
\citep{Schwanhauer.1982}.

This formula is dependent on the share of passenger trains 
\begin{equation}
    p_{\text{pt}, r} = \frac{\text{number of passenger trains on route $r$}}{\text{total number of trains  on route $r$}}
\end{equation}
on route $r$.
The more passenger trains are in the operating program of this route, the higher the threshold for sufficient timetable quality.
Some intuition into this can be obtained by looking at allowances for time deviations in the timetabling process in Germany \citep{dbinfragoNutzungsbedingungenNetzDB2023}.
While the timetable constructor might only move passenger trains up to $3 ~ \text{min}$, freight trains may be shifted by up to $30 ~ \text{min}$ - depending on the ordered type.

\subsection{Approximation Formulas}
\label{sec:approx_formel}

Different techniques have been developed to approximate the general independent queueing system (GI/GI).
In this section, we focus on two well-known approximation formulas in \citet{hertel1984exakte} (Section \ref{subsubsec:Hertel}) and \citet{kingmanSingleServerQueue1961} (Section \ref{subsubsec:Kingman}), while the utilizing of phase-type distributions in the Markov Chain to model general independent stochastic processes is presented in Section \ref{subsec:phase_type}.

\subsubsection{Hertel Approximation}
\label{subsubsec:Hertel}

To achieve the coefficients of variation of $v_A = 0.8$ and $v_B = 0.3$, 
the calculated queue-lengths can be scaled with the approximation formula 
\begin{equation}
\label{approx_ELW}
    L_{r}(M/M) \cdot \frac{1}{\gamma} \approx L_{r}(GI/GI)
\end{equation}
from \citet{hertel1984exakte} (see also \citep{fischer1990bedienungsprozesse}).

For this, the parameters
\begin{equation}
\label{approx_ELW_gamma}
    \gamma = \frac{2}{c \cdot v_B^2 + v_A^2}
\end{equation}
and
\begin{equation}
\label{approx_ELW_c}
    c = \left(\frac{\rho_r}{s}\right)^{1-v_A^2} \cdot (1+v_A^2) -v_A^2.
\end{equation}
are determined in dependence on the occupation ration $\rho_r=\lambda_r / \mu_r$, the amount of channels $s=1$ and the coefficients of variation $v_A, v_B$.
This has been used in \citet{emundsEvaluatingRailwayJunction2024} to model non-exponential arrival and service time distributions.

\subsubsection{Kingman Approximation}
\label{subsubsec:Kingman}

Another Approximation, that has been widely used for single-channel systems (see f.e. \citet{gudehusStaueffekteVorTransportknoten1976a}), is the Kingman approximation formula \citep{kingmanSingleServerQueue1961}
\begin{equation}
\label{approx_ELW_kingman}
    L_{r}(M/M) \cdot \left( \frac{v_A^2 + v_B^2}{2}\right) \approx L_{r}(GI/GI).
\end{equation}
It gives a good approximation for single-channel scenarios with a high occupation ratio $\rho \rightarrow 1$.

\subsection{Phase-Type Distributions}
\label{subsec:phase_type}

To use the estimated queue lengths $EL_r$ with the Markovian model, they are scaled with one of the approximation formulas in Section \ref{sec:approx_formel} to facilitate arbitrary coefficients of variation for the interarrival and service distributions.

In this section, we introduce a novel Continuous-Time Markov Chain formulation, enabling a direct modelling of general independent distributions.

For this, phase-type distributions \citep{cox1955use}, a type of distribution function, whose parameters can be fitted to approximate any other arbitrary probability distribution \citep{asmussen2003applied}, can be used.
Additionally, they can be represented by a Continuous-Time Markov chain, which facilitates their integration into the described model.

Specifically, the service and arrival processes on railways can be modelled with hypoexponential distributions because their assumed coefficients of variation $v_X < 1,~ X \in \{A, B\}$ are less than 1 (cf. \cite{Weik.2020PhD}).

\begin{figure}[htp]
    \centering
    \begin{subfigure}{.2\textwidth}
        \centering
        \resizebox{!}{2.9\baselineskip}{\input{Media/Markov-Graphs/exponential_graph}}
        \caption{M}
        \label{fig:Markov_service}
    \end{subfigure}%
    \begin{subfigure}{.8\textwidth}
        \centering
        \resizebox{!}{2.9\baselineskip}{\input{Media/Markov-Graphs/phase_type_graph}}
        \caption{Ph}
        \label{fig:phase_type_service}
    \end{subfigure}%
    \caption{Examples of the service process with exponential (M) or phase-distributed (Ph) service times}
    \label{fig:service_distribution}
\end{figure}
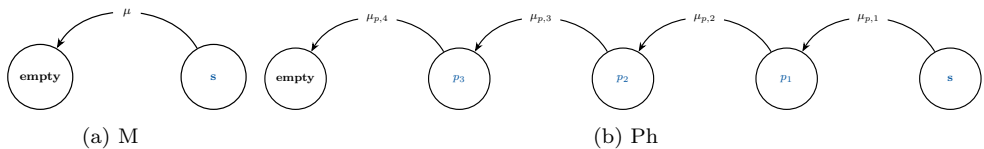

Figure \ref{fig:service_distribution} illustrates two different types of service processes: one with exponentially distributed service times (Figure \ref{fig:Markov_service}) and one with hypoexponentially distributed service times (Figure \ref{fig:phase_type_service}). In the Markov chain for the latter distribution, three intermediate states were introduced, which can be reached through transitions with different rates from the preceding states.

The number $k$ of the minimum required phases to approximate a distribution $X$ with coefficient of variation $v_X$ can be determined using
\begin{equation}
\label{k_phases_determination}
   k \geq \lceil \frac{1}{{v_X}^2} \rceil ~,
\end{equation}
(see \citet{david1987least}).

\subsubsection{Hypoexponential Parameter Fitting}
\label{subsubsec:hypo_para_fitting}

To adapt the hypoexponential distribution $\hat{X}$ to a distribution $X$, both the number of phases $k$ and the transition rates $\mu_{p_i}$ of the corresponding phases can be used. \citet{Weik.2020PhD} adapts an approach by \citet{sommereder2011modelling}, which divides the phases into two successive Erlang distributions $X_1, X_2$. The $k$ phases are determined using the coefficient of variation according to (\ref{k_phases_determination}) and divided into $k_{X_1} = \lceil \frac{k}{2} \rceil$ and $k_{X_2} = k - k_{X_1}$ phases.

In a first step, a combined distribution $(X_1^{\ast}, X_2^{\ast})$ of two artificial Erlang distributions with $k_{X_1}$/$k_{X_2}$ phases is fitted to the coefficient of variation $v_X$.
For this, the expected value of $X_1^{\ast}$ is fixed to $E(X_1^{\ast}) = 1$.  

By determining the expected value 
\begin{equation}
\label{calc:exp_x2_star}
    E({X_2}^{\ast}) = \frac{k_{X_1} k_{X_2} {v_X}^2 + \sqrt{k_{X_1} k_{X_2} ({v_X}^2 (k_{X_1} + k_{X_2} ) -1 )}}{k_{X_1} (1 - {v_X}^2 k_{X_2})}
\end{equation}
for the artificial Erlang distribution ${X_2}^{\ast}$, the target coefficient of variation 
\begin{equation}
    v_X = \frac{\sqrt{Var({X_1}^{\ast})+ Var({X_2}^{\ast})}}{E({X_1}^{\ast})+E({X_2}^{\ast})} = \frac{\sqrt{1/k_{X_1} + E({X_2}^{\ast})^2/k_{X_2}}}{1+E({X_2}^{\ast})}
\end{equation}

can be approximated together with the artificial Erlang distribution ${X_1}^{\ast}$.

To correctly estimate the expected value $E(X)$ of the distribution $X$, the expected values of the artificial Erlang distributions are then adjusted via
\begin{equation}
\label{calc:exp_x1}
E(X_1) = \frac{E(X)}{1+E({X_2}^{\ast})}
\end{equation}
and
\begin{equation}
\label{calc:exp_x2}
E(X_2) = \frac{E(X)E({X_2}^{\ast})}{1+E({X_2}^{\ast})}.
\end{equation}

Finally, the transition rate $\mu_{p_i}$ between phases $i-1$ and $i$ can then be determined by
\begin{equation}
\label{calc:mu_pi}
\mu_{p_i} =
    \begin{cases} 
      \frac{k_{X_1}}{E(X_1)} & i \leq k_{X1} \\
      \frac{k_{X_2}}{E(X_2)}  & i > k_{X1} ~.
   \end{cases}
\end{equation}

\subsubsection{Example}

In the example from Figure \ref{fig:phase_type_service}, the following parameters would be used to define a corresponding hypoexponential distribution for a service process with coefficient of variation $v_B = 0.5$ and a mean service time of $\bar{b}=3$: 
In total,
\begin{equation}
k = 4 = \frac{1}{{0.5}^2}
\end{equation}
phases need to be considered, with $k_{X_1} = k_{X_2}= 2$ states for the two artificial Erlang distributions.
Using \eqref{calc:exp_x2_star} - \eqref{calc:mu_pi}, it follows that
\begin{equation}
\mu_{p_i} = 
\begin{cases}
1.33 & 1 \leq i \leq 2\\
1.33 & 2 < i \leq 4 
\end{cases}
\end{equation}
for the transition rates between the states of the 4 phases, where the first transition represents the transition from state $\textcolor{blue}{\textbf{s}}$ to state $\textcolor{blue}{\textbf{$p_1$}}$, and the last transition represents the transition from state $\textcolor{blue}{\textbf{$p_3$}}$ to state $\textbf{empty}$.
Note that the states may have additional specifications, such as the number of waiting requests, which are omitted here.

\subsubsection{States}

To model hypoexponential distributions for the service and/or arrival processes, the current phase of the service and/or arrival processes must be added to the state description for each route.
A comprehensive definition of the state space $\hat{S}_{M,M}$ for purely exponentially distributed inter-arrival and service times can be found in \citet{emundsEvaluatingRailwayJunction2024}.

The corresponding state set
\begin{equation}
\label{zustandsmenge_m-m}
    \hat{S}_{M,M} = \left\{\left(q_1, s_1, \dots, q_k, s_k\right) | q_r \in \{0, \dots, m\}, s_r \in \{0,1\} \right\}
\end{equation}
can be extended to the state set
\begin{equation}
\label{zustandsmenge_ph-ph}
    \hat{S}_{Ph,Ph} = \left\{ \left(q_1, s_1, p_{A,1}, p_{S,1} \dots, q_k, s_k, p_{A,k}, p_{S,k}\right) |
    \left(q_r, s_r, p_{A,r}, p_{S,r}\right) \in \tilde{S}_r \right\}
\end{equation}
where with 
\begin{equation}
    \tilde{S}_r = 	\{0,\dots,m \}	\times \{0,1 \} \times \{0,\dots,k_{A, r}-1  \}	\times \{0,\dots,k_{S, r}-1  \}
\end{equation}
the set of tuples can be specified that describe the state of this track for a route $r$.

Thus, the number of waiting requests $q_r \in \{0,\dots,m\}$, the state of use of the track $s_r \in \{0, 1\}$, as well as the current phase of arrival $p_{A,r} \in \{0,\dots,k_{A, r}-1\}$ and service process $p_{S,r} \in \{0,\dots,k_{S, r}-1\}$ are described for each route $r$.
Here, $k_{A, r}$ and $k_{S, r}$ denote the number of phases in the arrival or service process.

For readability, the notations $q_r(u)$ for the queue-length $q_r$, $s_r(u)$ for the service status $s_r$, and $p_{A,r}(u), p_{S,r}(u)$ for the phase of the arrival/service process $p_{A,r}, p_{S,r}$ of route $r$ at state $u$ are additionally used in the following.

Furthermore, the states-space $\hat{S}_{Ph,Ph}$ can be restricted to
\begin{equation}
    S_{Ph,Ph} = \left\{u \in \hat{S} | \sum_{i=1}^k \sum_{\substack{j=1 \\ j\neq i}}^{k} \left(C_{i,j} \cdot s_i(u) \cdot s_j(j)\right) = 0 \right\}
\end{equation}
by using the conflict matrix $C$ to exclude states $u^{\ast} \in \hat{S}_{Ph,Ph} \setminus S_{Ph,Ph} $, that can not be visited because of conflicting ($C_{i,j} = 1$) routes, being serviced at the same time $s_i(u^{\ast}) = s_i(u^{\ast})=1$.

By formulating all transitions $T$, the railway junction can be modelled as a Continuous-Time Markov chain $MC = \left( S_{Ph,Ph}, T \right)$.
To map the hypoexponential distribution in the arrival or service process, the arrival or service transitions in \citep{emundsEvaluatingRailwayJunction2024} are replaced by transitions between individual phases in the respective process.

\subsubsection{Transitions}
\label{subsubsec:transitions}

Different transition types can be formulated on the set of states $S = S_{Ph,Ph}$. 
\textit{Arrival} transitions are introduced for the arrival of a train, \textit{service} transitions describe the service process and \textit{choice} transitions are used to decide between the next route service, if multiple are possible.

The \textbf{arrival} process to a route $r$ has been described by a transition $t=(u,v)$ with $q_r(v)= q_r(u)+1$ in the M/M model.
However, for the use of a phasetype distribution with a total of $k_{A,r}$ phases, a change in rates after $k_{A,r}^{\ast}$ phases and the two rates $\lambda_{r,a}$ and $\lambda_{r,b}$, multiple transitions $t=(u,v)$ need to be formulated for the arrival of a train.
They can be distinguished according to the arrival phase $p_{A,r}(u)$ from the start state of a transition.
For $p_{A,r}(u) \leq k_{A,r}^{\ast}$, transitions from $u$ to $v$ with
\begin{equation}
    p_{A,r}(v) = p_{A,r}(u) + 1, 
\end{equation}
are applied with a rate of $\lambda_{r,a}$.
Furthermore, for $k_{A,r}^{\ast} \leq p_{A,r}(u) \leq k_{A,r}$,  transitions of
\begin{equation}
    p_{A,r}(v) = p_{A,r}(u) + 1, 
\end{equation}
with rate $\lambda_{r,b}$ are added.
Lastly, transitions starting the service or adding the train to the queue are formulated for $p_{A,r}(u) = k_{A,r}$.
When no conflicting routes are serviced in the start state $u$, the service of route $r$ can start in state $v$
\begin{equation}
    s_{r}(v) = 1, 
\end{equation}
otherwise a train is added to the queue
\begin{equation}
    q_{r}(v) = q_{r}(u) + 1 
\end{equation}
of the route.
Both possible transitions reset the phase of the arrival process to
\begin{equation}
    p_{A,r}(v) =  1
\end{equation}
in the end state $v$ and use a rate of $\lambda_{r,b}$.

\textbf{Service} transitions between states $u,v$, with $s_{r}=1$, are similarly modelled.
Let $k_{S,r}$ be the number of phases in the service process with a change from rate $\mu_{r, a}$ to $\mu_{r, b}$ after $k_{S,r}^{\ast}$ phases.
Then, for $p_{S,r}(u) \leq k_{S,r}^{\ast}$ transitions 
\begin{equation}
\label{transition_service_+1}
    p_{S,r}(v) = p_{S,r}(u) + 1, 
\end{equation}
with a rate of $\mu_{r, a}$ are used.
For $ k_{S,r}^{\ast} \leq p_{S,r}(u) < k_{S,r}$, the rate changes to $\mu_{r, b}$ and the same property \eqref{transition_service_+1} for $u$ and $v$.
Finally, if $p_{S,r}(u) = k_{S,r}$, the service process is terminated by resetting
\begin{equation}
    p_{S,r}(v) =  1.
\end{equation}

The last type, \textbf{choice} transitions, are used to model the initiation of service for a route when multiple routes $r \in R$ can be serviced due to the absence of conflicting routes being active in the initial state $u$.
For each such possible route $r^{\ast}$, a transition $t = (u, v)$ with an artificial rate $M$ is introduced, ensuring that
\begin{equation}
    s_{r^{\ast}}(v) = 1
\end{equation}
and
\begin{equation}
    q_{r^{\ast}}(v) = q_{r^{\ast}}(u) - 1 
\end{equation}
are satisfied.

Ideally, these choice transitions would not introduce any additional time into the system. 
Therefore, the rate $M$ should be selected to be sufficiently high, such that the expected induced time $1/M$ remains negligibly small. 
For all models presented in this paper, a rate of $M = 600$ has been selected, corresponding to a delay of $1/600$ minutes, which is considered to provide an adequately precise approximation (see also \citep{emundsEvaluatingRailwayJunction2024}).

Additionally, transitions from other processes must also be incorporated into the introduced intermediate states.
Consequently, the arrival of a request on a route $i^{\prime}$ can be modeled during phase $j$ of the service process for another route $i$.
A full description of a model in the PRISM modelling language can be found in the online repository \citep{emundsResearchDataCode2024}.

The modeling approach introduced here, which employs hypoexponential distributions for the arrival and/or service processes, is evaluated for accuracy in Section \ref{sec:validation}.

\subsection{Queue-Length Analysis}
\label{subsec: Queue-Length Analysis}

The introduced Continuous-Time Markov Chain $MC=\left(S_{Ph,Ph},T\right) = \left(S, T\right)$ models the process of scheduling train requests on the considered railway junction.
In order to assess the performance of this junction, the length $L_{r}$  of the queue can be determined for all routes $r$ by analysing the probability $p(u)$ of the states $u \in S$ with certain queue-lengths in the stationary distribution, i.e. the probability that the system is in state $u$ in the long run.
Hence, the expected length of the queue can be calculated with
\begin{equation}
    L_{r} = \sum_{\substack{u \in S \\ q_r(u) > 0}} p(u) \cdot q_r(u).
\end{equation}

The same basic concept has already been used by other analytical approaches in the performance determination of railway infrastructure \citep{Schmitz.2017, Weik.2017, Weik.2020PhD, emundsEvaluatingRailwayJunction2024}.
While various solution methods to determine the state probabilities $p(u)$ have been used, \citet{emundsEvaluatingRailwayJunction2024} introduced an approach that builds the CTMC in the formal PRISM language \citep{DaveParkerGethinNormanMartaKwiatkowska.2000} and applies probabilistic model-checking \citep{Hensel.2022} to obtain the expected queue-lengths $L_{r}$.
The same mechanism is used for this work.

\subsection{Capacity Determination Algorithm}
\label{subsec:cap_deter_alg}

To determine the timetable capacity of a railway junction, multiple $n_{\text{total}}$ values must be tested to find the maximum train count $n_{\text{max}}$, where the expected queue lengths $L_r$ do not exceed the limit $L_{\text{limit},r}$ (Section \ref{subsec:threshold}).
With the introduced approach of modeling phase-type distributions directly within the CTMC, the number of states increases significantly (see also Section \ref{sec:validation}), leading to considerably longer computation times for the expected queue lengths $L_r$.
It is therefore crucial to minimize the number of iterations in which the queue lengths must be recalculated, i.e., the number of tested $n_{\text{total}}$ values.

To achieve this, a root finding problem can be formulated.
Let the function
\begin{equation}
     EL_{r}: N_{\text{total}} \rightarrow \mathbb{R},  n_{\text{total}} \mapsto EL_{r} (n_{\text{total}})
\end{equation}
describe the estimation of the expected queue-length $EL_{r}(n_{\text{total}}) = L_r$ for a given train count $n_{\text{total}}\in N_{\text{total}}$ by utilizing probabilistic model-checking on a formulated CTMC.

By comparing it to the limit, the quality factor
\begin{equation}
    qf_r (n_{\text{total}}) = \frac{EL_{r}(n_{\text{total}})}{L_{\text{limit}}}
\end{equation}
of a route $r$ can be obtained. 
Since the limit has to be adhered to for every route, the optimal $n_{\text{total}}^* = n_{\text{max}}$ fulfills the property
\begin{equation}
    qf_{\text{max}} \left(n_{\text{total}}^*\right) = \max_{r \in R} qf_r \left(n_{\text{total}}^*\right) = 1,
\end{equation}
such that the maximum quality factor $qf_{\text{max}}$ is exactly $1$, and therefore the queue length at one route is exactly at the limit for this route.

The function
\begin{equation}
    \phi: N_{\text{total}} \rightarrow \mathbb{R}: n_{\text{total}} \mapsto \phi\left(n_{\text{total}}\right) = qf_{\text{max}} \left(n_{\text{total}}\right) - 1
\end{equation}
can now be formulated, whose root $\phi\left(n_{\text{total}}^*\right)=0$ describes the timetable capacity $n_{\text{total}}^* = n_{\text{max}}$ of the railway junction.

To determine the root of $\phi$, the algorithm from \citet{brent1973algorithms}, as implemented in \texttt{scipy} \citep{2020SciPy-NMeth}, was utilized.
This algorithm efficiently computes the root of a function $f:\mathbb{R} \rightarrow \mathbb{R}$ within a specified interval $[a, b]$, where $f$ changes its sign, achieving a solution up to a chosen level of accuracy.
In the validation implementation (Section \ref{subsec:perf_det_quality}) and the case study (Section \ref{sec:case_study}), a termination criterion was set to ensure both a total error of less than $10^{-3}$ and a relative error below $10^{-3}$.
An Algorithm, describing how to obtain the function evaluation $\phi(n_{\text{total}})$ for a given total train count $n_{\text{total}}$ is given in \ref{app:cap_deter_algo}.

This algorithm is then called consecutively from within Bent's method for varying $n_{\text{total}}$, until the set tolerances are met and $n_{\text{max}}$ has been found.

\section{Validation}
\label{sec:validation}

To analyze the different distribution functions, the following distinguishes between four different applications.

\begin{itemize}
    \item Exponentially distributed inter-arrival and service times (M/M)
    \item Phase-type distributed inter-arrival and exponentially distributed service times (PH/M)
    \item Exponentially distributed inter-arrival and phase-type distributed service times (M/PH)
    \item Phase-type distributed inter-arrival and service times (PH/PH)
\end{itemize}

For each distribution of arrival and service processes, a model of the junction from Figure \ref{fig:junction} was created.

\begin{table}[th]
\caption{Parameters of the studied models}
\label{table:model_paras}
\resizebox{\textwidth}{!}{
\begin{tabular}{c|c||c|c|c|c|c|c|c}
\thead{Arrival \\ process} & \thead{Service \\ process} & \thead{Number \\ of phases \\ Arrival} & \thead{Number \\ of phases \\ Service} & \thead{$v_A$} & \thead{$v_S$} & \thead{Number \\ of waiting slots \\ $m$} & \thead{Number \\ of states} & \thead{Number \\ of transitions} \\ \hline \hline
M  & M  & 1 & 1 & 1   & 1   & 5 & 10 368    & 63 688     \\ \hline
Ph & M  & 2  & 1 & 0.8 & 1   & 5 & 141 108   & 829 825    \\ \hline
M  & Ph & 1 & 12  & 1   & 0.3 & 5 & 623 376   & 3 664 703  \\ \hline
Ph & Ph & 2  & 12  & 0.8 & 0.3 & 5 & 8 192 448 & 46 447 056
\end{tabular}
}
\end{table}

In Table \ref{table:model_paras}, the parameters of the different models are presented.
Particularly noteworthy is the influence of the hypoexponential distribution on the size of the model.
While the number of states for the model with exponentially distributed inter-arrival and service times is still in the order of $10^{4}$, more than 8 million states must be considered for the model with hypoexponentially distributed inter-arrival and service times.

The solving process has been automated using Python 3.10.9 \citep{van1995python, PythonSoftwareFoundation.2022} and the probabilistic model-checker \texttt{Storm} \citep{Hensel.2022} with its Python-Interface \texttt{stormpy} \citep{SebastianJunges.2023}.
Furthermore, \texttt{scipy} \citep{2020SciPy-NMeth} has been used for an implementation of the Brent’s method (see Section \ref{subsec:cap_deter_alg}).
Used CTMC models (see Section \ref{subsec:phase_type}) have been expressed in the PRISM modelling language \citep{DaveParkerGethinNormanMartaKwiatkowska.2000, KNP11}.
The used model files can be found in the online repository \citep{emundsResearchDataCode2024}.

Results are compared to simulations, that have been implemented in Python 3.10.9 \citep{van1995python, PythonSoftwareFoundation.2022}, utilizing \texttt{SimPy} \citep{TeamSimPyRevision.2023} and \texttt{Ciw} \citep{ciwpython, ciwarticle}.
An implementation can be found in the online repository \citep{emundsResearchDataCode2024}.
In the simulation, inter-arrival times and service times are drawn randomly from phase-type distributions fitted according to Section \ref{subsubsec:hypo_para_fitting} for every route.
Trains are managed using first-in-first-out (FIFO) queues for every route, and conflicting routes are prevented from starting service simultaneously by ensuring that shared resources are not utilized concurrently. 
The validity of this approach is verified through conflict analysis and the queue-length per route is monitored at each simulated minute and averaged for comparison purposes.

In the following sections, computation times and the approximation quality of the different experiments are analysed.

\subsection{Computation Time}

\begin{figure}[ht]
    \centering \includegraphics[width=.8 \textwidth]{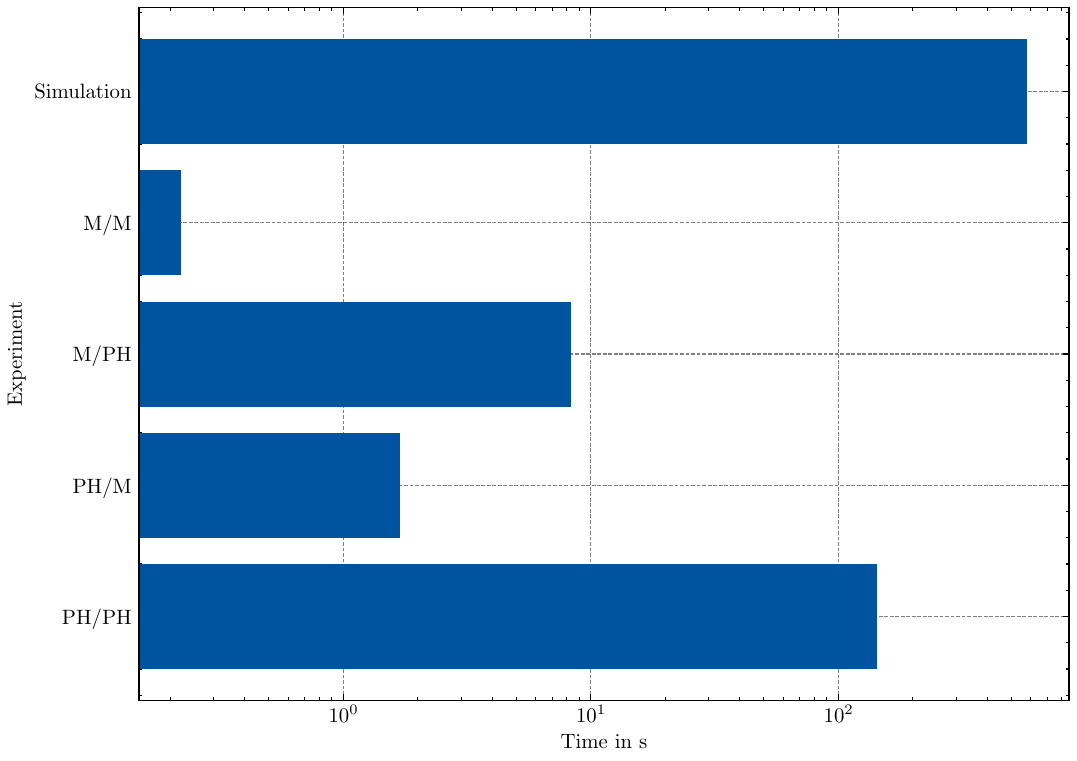}
    \caption{Mean computation times of all considered models}
    \label{fig:computation_time_bars}
\end{figure}

All experiments have been conducted on an Intel Xeon Platinum 8160 Processor (2.1 GHz) and with a working memory of 32 GB. Additionally, 100 Simulations with 20 hours each have been performed on the same processor with 4 GB of RAM for every considered train count $n_{total}$.

When examining the computational times necessary to determine the queue lengths in the individual experiments, the size of the state-space becomes a significant factor.
Figure \ref{fig:computation_time_bars} shows the median of the respective calculations for each model.
Furthermore, Figure \ref{fig:computation_time_bars} includes the computational time required to simulate $2 000$ hours, which does not differ significantly between the assumed distributions of inter-arrival and service times.
It is evident that the computation time increases substantially with the number of states.

Using the model for exponentially distributed inter-arrival and service times, approximately $0.22$ seconds are required to compute the expected queue length in the median.
In contrast, computing the expected queue length using the model with arrival and service processes following a hypoexponential distribution requires over $140$ seconds.
However, all analytical methods can calculate the queue length faster than a simulation of $2 000$ hours, which takes a median time of over $550$ seconds.

\subsection{Queue-Length Approximation Quality}

In addition to the computation times of the different models, the approximated queue-lengths can be compared.
For this, an exemplary railway junction with 4 routes (see Figure \ref{fig:junction}) has been chosen.
For this example, we assume passenger traffic only, hence resulting in a quality threshold (see Section \ref{subsec:threshold}) of $L_{limit} = 0.479 \cdot \text{exp}(-1.3) = 0.13$.
The service rates have been assumed to have artificial values of $\mu_r= 0.3$ for every route $r \in R$, corresponding to a mean minimum headway time of $\bar{b}= 1/0.3 \approx 3.33 ~ \text{min}$.
To highlight the dependencies of the results on the occupancy rate $\rho$, the total number of trains in the infrastructure per hour $n_{\text{total}} = \sum_{r \in R} n_{r}$ has been varied from $n_{\text{total}} \in \{4, 8, \dots, 40\}$ by setting the arrival rate $\lambda_r = n_{r}/60$ for every route $r \in R$. 
For a fixed traffic distribution of $p_{\text{main}} = 0.5$, the number of trains per route $n_r$ is given by $n_r = n_{\text{total}}/4$ for every route.
The arrival-rate per route $\lambda_r$ has hence been set to values between $1/60$ and $10/60$.

The expected queue length $EL_r$ for each route $r$ has been calculated for every experimental distribution setting.
Furthermore, for the models with exponential arrival and/or service processes, the approximation formulas by Hertel (Section \ref{subsubsec:Hertel}) and Kingman (Section \ref{subsubsec:Kingman}) have been applied to approximate the expected queue-length of the general independent system $EL_r(GI/GI)$.
In their formulas, the coefficients of variation $v_A$ and $v_B$ have been set according to the used distribution setting:
If a coefficient of variation $v_X \neq 1$ has been used in the model, the corresponding coefficient of variation has been set to $v_X = 1$ in the approximation formula.
Otherwise a coefficient of variation of $v_A = 0.8$ has been used for the arrival process and of $v_S = 0.3$ for the service process.
Table \ref{table:model_paras_approx} lists the values in the approximation formulas.

\begin{table}[h]
\centering
\footnotesize
\caption{Parameters of the approximation settings}
\label{table:model_paras_approx}
\resizebox{1\textwidth}{!}{
\begin{tabular}{c|c||c|c|c|c|c|c}
\thead{Arrival \\ process} & \thead{Service \\ process} & \thead{approx. \\ formula \\ $v_A$} & \thead{approx. \\ formula \\ $v_S$} & \thead{service rate \\ per route \\ $\mu_r$} & \thead{total \\ train count  \\ $n_{total}$} & \thead{quality \\ threshold \\$L_{limit}$} & \thead{share of \\ main traffic \\ $p_{main}$} \\ \hline \hline
M  & M   & 0.8   & 0.3   & 0.3 & $4, \dots, 40$ & $0.13$ & $0.5$       \\ \hline
Ph & M   & 1 & 0.3   & 0.3 & $4, \dots, 40$ & $0.13$  & $0.5$    \\ \hline
M  & Ph   & 0.8   & 1 & 0.3 & $4, \dots, 40$ & $0.13$  & $0.5$    \\ \hline
Ph & Ph  & 1 & 1 & 0.3 & $4, \dots, 40$ & $0.13$ & $0.5$
\end{tabular}
}
\end{table}

\begin{figure}[h!]
\centering
\begin{subfigure}{.5 \textwidth}
    \centering
    \includegraphics[width=\textwidth]{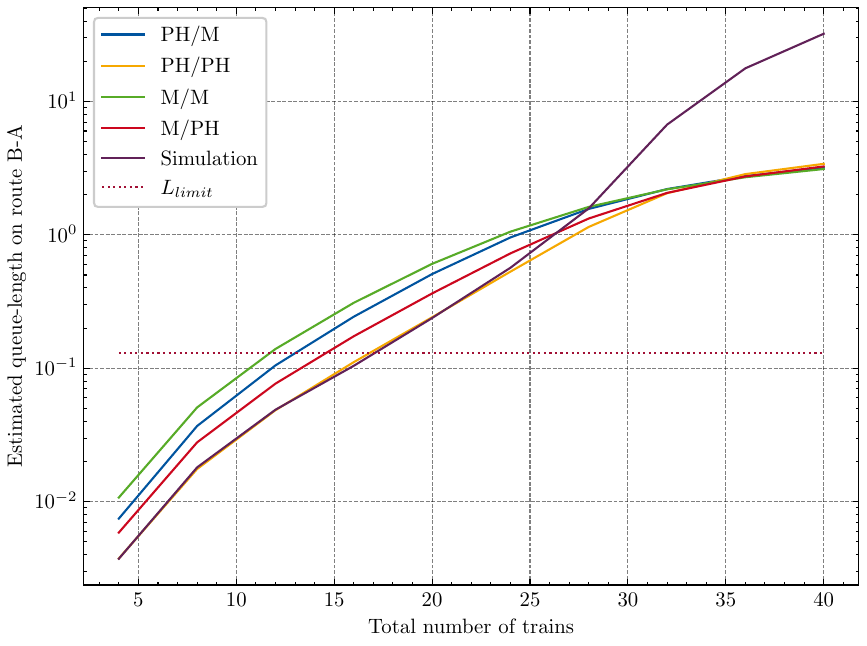}
    \caption{Without Scaling Factor}
    \label{fig:wsl_per_n_tot_no_scaling}
\end{subfigure}\vspace{0.5cm}
\begin{subfigure}{.5\textwidth}
    \centering
    \includegraphics[width=\textwidth]{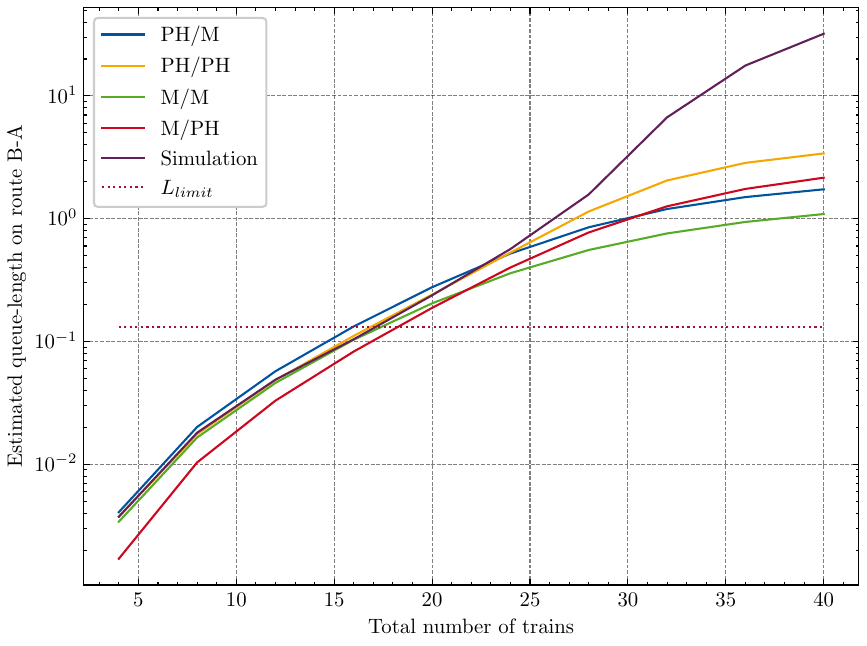}
    \caption{Hertel Approximation}
    \label{fig:wsl_per_n_tot_hertel}
\end{subfigure}%
\begin{subfigure}{.5\textwidth}
    \centering
    \includegraphics[width=\textwidth]{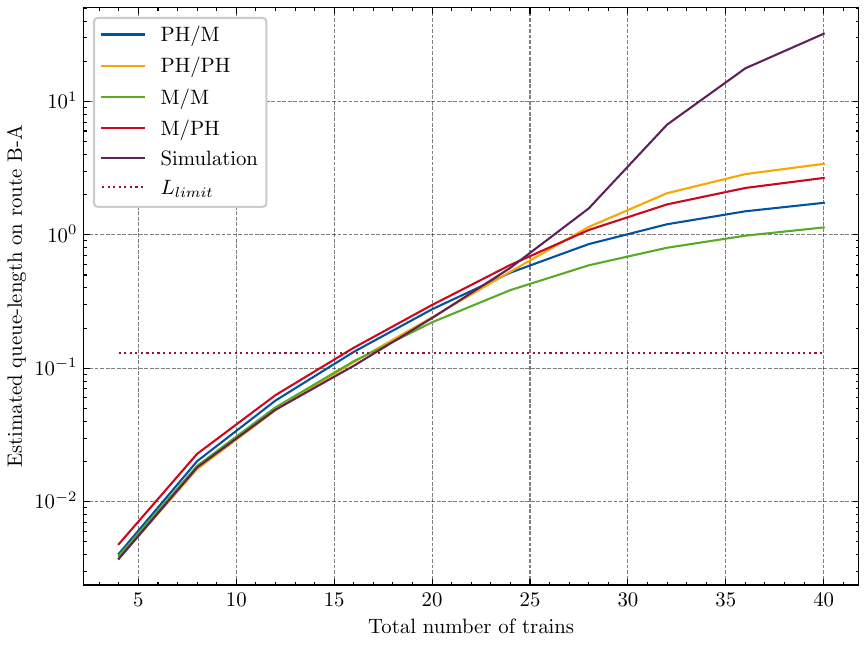}
    \caption{Kingman Approximation}
    \label{fig:wsl_per_n_tot_kingman}
\end{subfigure}
\caption[short]{Comparison of the queue-length estimations at route $r_3$ for different arrival- and service-process distributions with and without scaling factors}
\label{fig:wsl_per_n_tot}
\end{figure}

To analyse the approximation quality of a single route, a fixed traffic distribution of $p_{\text{main}}= 0.5$ has been set for varying $n_{\text{total}}$.
Selecting route $r_3$ from B to A, Figure \ref{fig:wsl_per_n_tot} shows the estimated values for the queue-length at this route and compares them to the mean of the simulation.
The values have been calculated with the specified model and either not scaled at all (Figure \ref{fig:wsl_per_n_tot_no_scaling}) or approximated with the Hertel (Figure \ref{fig:wsl_per_n_tot_hertel}) or Kingman (Figure \ref{fig:wsl_per_n_tot_kingman}) method.
In this example, all four routes $r \in R$ are assumed to have the same number of trains $n_r = n_{total}/4$ in an hour.

In all three Figures \ref{fig:wsl_per_n_tot_no_scaling} - \ref{fig:wsl_per_n_tot_kingman}, the graphs of the PH/PH model and the simulation are almost congruent for a train count $n_{total}$ below $23$.
At higher train numbers, the difference between the two calculates queue-lengths increases.
This can be explained with the limitations in the maximum queue-length, or the number of waiting slots $m$ in the queueing system, which has been set to $m=5$ in the analytical experiments, but remained unlimited for the simulation setup.
The probability of more than $5$ trains having to wait for their service increases with an increasing number of trains.
Consequently, the analytical models, which omit the state probabilities with more than $5$ simultaneous waiting trains, underestimate the total queue-length.
However, since the quality threshold is defined for a limit of $L_{limit} = 0.13$, the PH/PH model gives an adequate estimation in the neighborhood of the limit and can therefore be used without any scaling functions to evaluate the performance of the example railway junction.

For the graph (Figure \ref{fig:wsl_per_n_tot_no_scaling}) without any applied scaling, it is clearly evident that there are significant differences between the simulation method and the models with exponential arrival and/or service processes.
In these three experiments,  only $12-14.5$ trains per hour would be permitted to meet the given quality threshold specified quality threshold, whereas both the simulation method and the PH/PH model would set this limit at approximately 17 trains per hour. 
Additionally, the M/PH model demonstrates greater alignment with the simulation and PH/PH model compared to the PH/M model, which still exhibits more accuracy than the basic M/M model.
These discrepancies can be attributed to variations in the coefficients of variation $v_A$ and $v_S$ from 1 — the M/PH model accurately represents the service process (characterized by a smaller coefficient of variation), while the PH/M model provides a more precise description of the arrival process (with $v_A$ closer to $1$).

However, when an approximation method is applied, the M/M model closely aligns with the simulation and the PH/PH method. In the case of the Hertel approximation (Figure \ref{fig:wsl_per_n_tot_hertel}), significant discrepancies become evident for a total train count of $n_{total} \geq 17$. Similarly, for the Kingman approximation (Figure \ref{fig:wsl_per_n_tot_kingman}), differences become apparent for $n_{total} \geq 19$.

The models for the PH/M and M/PH settings consistently overestimate or underestimate the simulation results within the relevant region where the total train count is $n_{total} \leq 22$.
In the case of the Hertel approximation, the PH/M model yields an overestimation while the M/PH model yields an underestimation.
Alternatively, for the Kingman approximation, both models produce overestimations, albeit more closely aligned with the simulation results.

\subsection{Performance Determination Quality}
\label{subsec:perf_det_quality}

While the results for the raw queue-length estimations have been analyzed in the previous section, this section focuses on the application of the introduced methods for performance assessment.
In certain regions, such as those far from the quality threshold, a small deviation from the correct result in the analyzed parameter may not significantly impact the overall approximation quality of performance indicators.

Therefore, we apply the capacity determination algorithm from Section \ref{subsec:cap_deter_alg} for all four analytical settings and both introduced approximation methodologies (see Section \ref{sec:approx_formel}), the parameter settings can be found in Tables \ref{table:model_paras} and \ref{table:model_paras_approx}.
Table \ref{table:solving_paras} summarizes the results for the different calculation scenarios.

\begin{table}[h]
\centering
\footnotesize
\caption{Selected indicators for the different methods}
\label{table:solving_paras}
\resizebox{1\textwidth}{!}{
\begin{tabular}{c||c|c||c|c|c|c}
\thead{Approximation \\ formula} & \thead{Arrival \\ process} & \thead{Service \\ process} & \thead{Mean \\ number of \\ iterations} & \thead{Mean total \\ computation \\ time} & \thead{Maximum \\ capacity \\ $\hat{n}_{max}$} & \thead{Traffic share \\ at max. \\ capacity} \\ \hline \hline
None    & M  & M  & $9$ & $2.22$ s & $11.70$ & $0.5$ \\ \hline
None    & PH & M  & $9$ & $17.04$ s & $12.97$ & $0.5$  \\ \hline
None    & M  & PH & $8.44$ & $78.86$ s & $14.53$ & $0.5$  \\ \hline
None    & PH & PH & $9$ & $1432.20$ s & $16.90$ & $0.1$  \\ \hline \hline
Kingman & M  & M  & $8.89$ & $2.35$ s & $16.80$ & $0.5$  \\ \hline
Kingman & PH & M  & $8$ & $15.76$ s & $15.91$ & $0.5$  \\ \hline
Kingman & M  & PH & $7.78$ & $74.65$ s & $15.55$ & $0.5$  \\ \hline \hline
Hertel  & M  & M  & $9.33$ & $2.31$ s & $17.29$ & $0.5$   \\ \hline
Hertel  & PH & M  & $8$ & $15.70$ s & $15.91$ & $0.5$   \\ \hline
Hertel  & M  & PH & $9.22$ & $87.33$ s & $18.17$ & $0.5$  
\end{tabular}
}
\end{table}

The algorithm from Section \ref{subsec:cap_deter_alg} assumes monotonicity for the queue-length function $EL_{r}: P_{\text{main}} \times N_{\text{total}} \rightarrow \mathbb{R}$, which cannot be guaranteed for the non-deterministic simulations.
Therefore, 100 simulations, each lasting 20 hours, are conducted for every combination of $p_{\text{main}} \in P_{\text{main}} =\{0.1, \dots, 0.9\}$ and $n_{\text{total}} \in N_{\text{total}} = \{12.00, 12.04, \dots, 19.92, 19.96\}$.
From this, the bounds $n_{\text{max, LB}}^{\ast}$ and $n_{\text{max, UB}}^{\ast}$ can be identified to limit the optimal train-count $n_{\text{max}}$ in the simulation that meets the quality thresholds (see \ref{app:cap_deter_sim}).

\begin{figure}[htp]
\centering
\begin{subfigure}{1 \textwidth}
    \centering
    \includegraphics[width=\textwidth]{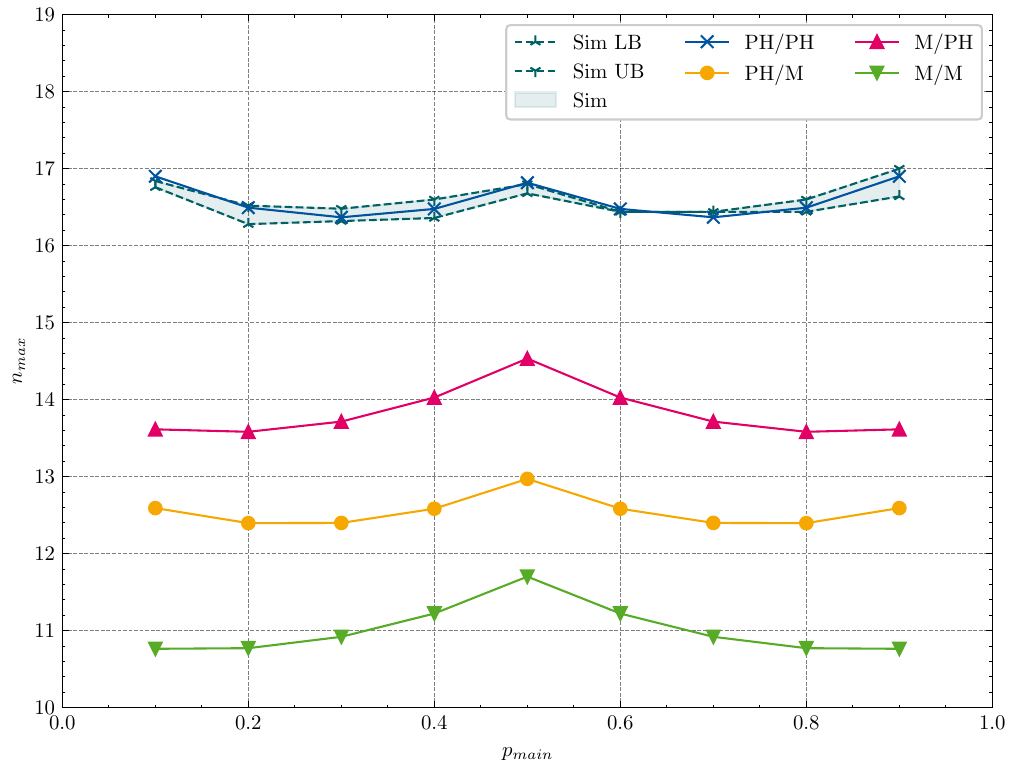}
    \caption{Without Scaling}
    \label{fig:perf_comp_no_scaling}
\end{subfigure}
\vspace{0.5cm}
\begin{subfigure}{.5\textwidth}
    \centering
    \includegraphics[width=\textwidth]{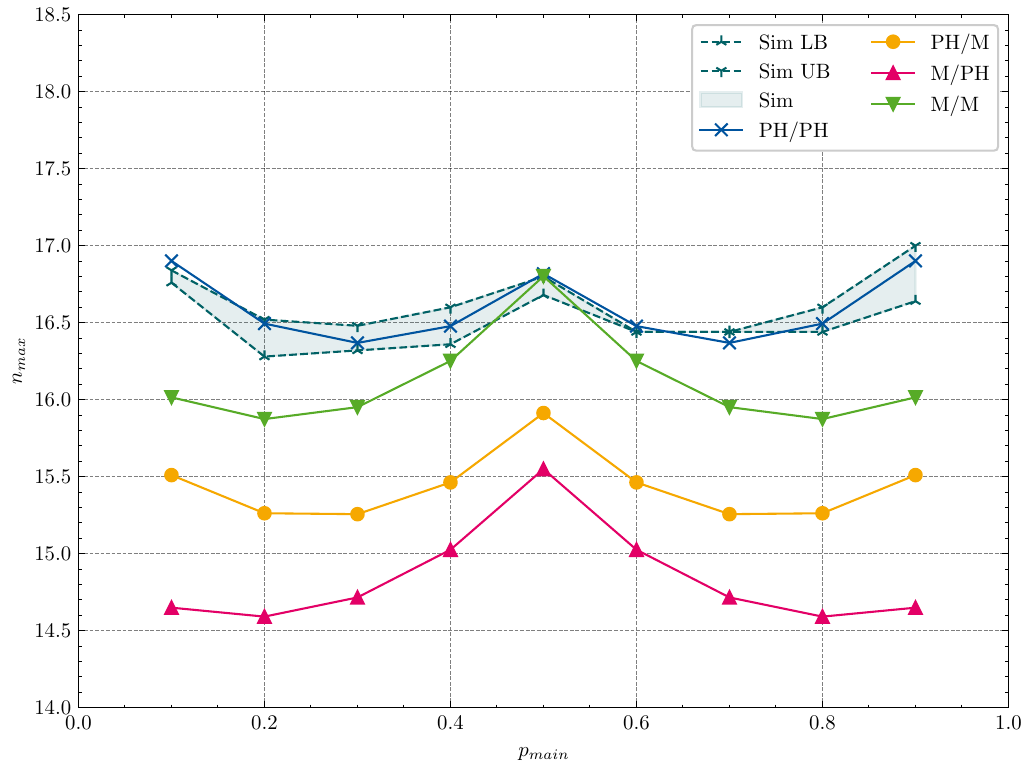}
    \caption{Kingman Approximation}
    \label{fig:perf_comp_kingman}
\end{subfigure}%
\begin{subfigure}{.5\textwidth}
    \centering
    \includegraphics[width=\textwidth]{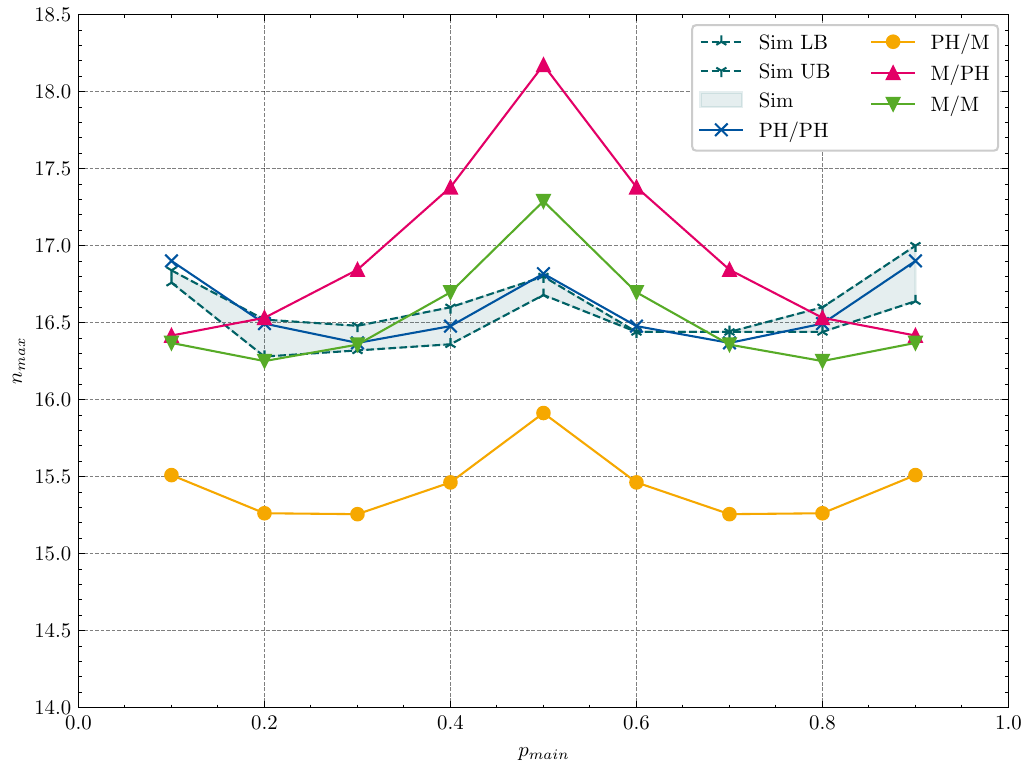}
    \caption{Hertel Approximation}
    \label{fig:perf_comp_hertel}
\end{subfigure}

\caption{Comparison of the performance determination by the share of main line traffic $p_{main}$ for the different methods}
\label{fig:perf_comp}
\end{figure}

In Figure \ref{fig:perf_comp}, the results of the performance determinations are depicted for the different methods.

When comparing the simulation results to the PH/PH method (Figure \ref{fig:perf_comp_no_scaling}), some minor deviations are apparent.
For example, the analytical solution exhibits a clear symmetry around the axis $p_{main} = 0.5$, which is only qualitatively reflected in the simulation results.
This demonstrates that even the averaged simulation results represent only a subset of all possible outcomes, whereas the analytical method performs the capacity determination deterministically.

Overall, the simulation and PH/PH results are similar in terms of capacity $n_{max}$ and exhibit the same general pattern: one local maximum for homogeneous traffic and two local maxima at the extremes, where nearly all traffic is concentrated on either the main or branch line.

For the raw analytical methods without any applied scaling factor (Figure \ref{fig:perf_comp_no_scaling}), the M/PH, PH/M, and M/M methods significantly underestimate the available capacity and fail to exhibit the same general pattern of two maxima at the extremes.

Applying the Kingman approximation (Figure \ref{fig:perf_comp_kingman}) significantly reduces the gap between the results of the M/M model and the PH/PH model, particularly for the homogeneous traffic scenario at $p_{main}=0.5$.
However, substantial differences in the calculated capacities persist in other traffic distributions.
Notably, the order of approximation accuracy changes with the Kingman approximation: the M/M model offers the best approximation, followed by the PH/M model, and lastly the M/PH model.

Using the Hertel formula (Figure \ref{fig:perf_comp_hertel}), the M/M model also provides the best fit to the PH/PH and simulation results.
Although the M/M model still fails to accurately capture the capacity increase for highly heterogeneous traffic distributions, its overall estimates are quite close, with differences of less than $0.5$ trains per hour at the local extrema of the PH/PH results.
The M/PH model overestimates capacity by up to $1.5$ trains per hour in the homogeneous scenario, intersects the PH/PH model near the boundaries, and underestimates performance at both the lowest and highest traffic shares.
In contrast, the PH/M model shows the exact same poor approximation quality as with the Kingman approximation, with a clear underestimation of capacity by at least $1$ to $1.5$ trains per hour.

In summary, the PH/PH model delivers the most accurate results, although it requires the highest computational time.
On the other hand, the Hertel approximation enables the M/M model to achieve a good fit for homogeneous traffic scenarios in a fraction of the time.
The use of PH/M or M/PH models is generally not advisable, as these models are more inaccurate after applying an approximation formula, while also demanding more computational time.
Depending on the application scenario, engineers may opt for the M/M model combined with the Hertel approximation for preliminary and rapid planning stages, while reserving the PH/PH model for more detailed analysis of junction infrastructure in selected long-term projects.

\section{Case Study}
\label{sec:case_study}

\begin{figure}[h!]
    \centering
    \input{case_study_overview.tikz}
    \caption{Traffic scenario of the case study}
    \label{fig:case_study_scenario}
\end{figure}
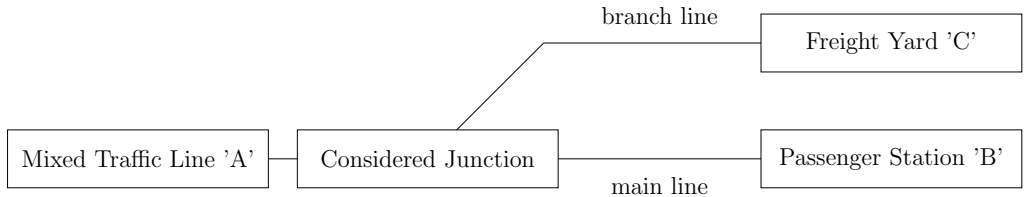

In this Section, we consider a double track railway junction, where a mixed-traffic railway line is divided into a freight line from and to a freight yard and a line for passenger traffic only (see Figure \ref{fig:case_study_scenario}).

\begin{figure}[h!]
    \centering
    \includegraphics[width=\textwidth]{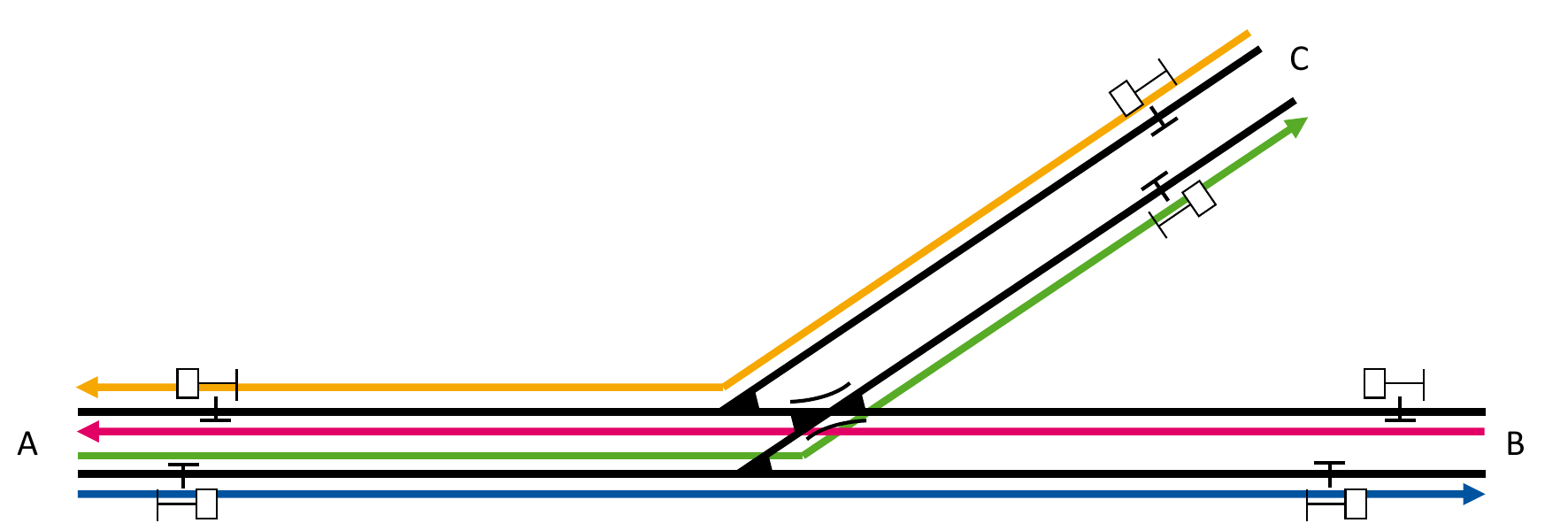}
    \caption{Infrastructure of the considered railway junction}
    \label{fig:junction_case_study}
\end{figure}

Consistent with the introduced notation, the line to the passenger station 'B' is called \textit{main line} and the line to the freight yard 'C' \textit{branch line}.
For the railway junction infrastructure itself, we consider the same infrastructure as in Figure \ref{fig:junction}, Figure \ref{fig:junction_case_study} redraws it for convenience.

\subsection{Setup}
\label{seubsection:case_study_setup}

The traffic on this infrastructure can be described with four different train types: \textit{Suburban} (s) and \textit{regional} (r) trains cover passenger traffic, while cargo trains can be distinguished between \textit{long-distance freight} (lf) and \textit{regional freight} (rf) trains.
Depending on their origin and destination, only certain routes are feasible for these types of trains (see Table \ref{tab:traffic_case_study}).

\begin{table}[h!]
    \centering
    \caption{Traffic in the case study}
    \label{tab:traffic_case_study}
    \begin{tabular}{c|c|c|c|c}
         \thead{Train types} & \thead{Route} & \thead{Origin} & \thead{Destination}  & \thead{Conflicts}\\ \hline \hline
         \makecell{suburban train, \\ regional train}& \textcolor{rwth}{$r_1$} & A & B & \textcolor{grun}{$r_2$} \\ \hline
         \makecell{long-dist. freight train, \\ regional freight train} & \textcolor{grun}{$r_2$} & A & C & \textcolor{rwth}{$r_1$}, \textcolor{magenta}{$r_3$}   \\ \hline
         \makecell{suburban train, \\ regional train} & \textcolor{magenta}{$r_3$} & B  & A & \textcolor{grun}{$r_2$}, \textcolor{orange}{$r_4$}  \\ \hline
         \makecell{long-dist. freight train, \\ regional freight train} & \textcolor{orange}{$r_4$}  & C & A  & \textcolor{magenta}{$r_3$}
    \end{tabular}
\end{table}

To assess the junction's performance, minimum headway times of the different routes have to be considered.
Usually, an infrastructure manager would utilise microscopic infrastructure data to calculate minimum headway times with their corresponding toolset.
However, for this made-up example, no microscopic data exists, therefore minimum headway times have been estimated in Table \ref{tab:headway_case_study}.

\begin{table}[h]
\centering
\caption{Assumed minimum headway times $h_{i,j}$ in minutes for the considered junction}
\label{tab:headway_case_study}
\resizebox{1\textwidth}{!}{
\begin{tabular}{|c||c|c|c|c|c|c|c|c|}
\hline
 $\left(r_i, t_i\right)$ \textbackslash ~ $\left(r_j, t_j\right)$ & $\left(r_1, s\right)$   & $\left(r_1, r\right)$   & $\left(r_3, s\right)$   & $\left(r_3, r\right)$   & $\left(r_2, lf\right)$   & $\left(r_2, rf\right)$  & $\left(r_4, lf\right)$   & $\left(r_4, rf\right)$   \\ \hline \hline
$\left(r_1, s\right)$ & 2.5 & 5.5 &  &  & 5 & 5 &  & \\ \hline
$\left(r_1, r\right)$ & 3 & 2 &  &  & 3 & 3 & & \\ \hline
$\left(r_3, s\right)$ & & & 2.5 & 5.5 & 1.5 & 1.5 & 5 & 5 \\ \hline
$\left(r_3, r\right)$ & & & 3 & 2 & 1.5 & 1.5 & 3 & 3 \\ \hline
$\left(r_2, lf\right)$ & 3.5 & 5.5 & 3 & 2 & 3 & 6 &  & \\ \hline
$\left(r_2, rf\right)$ & 8 & 8.5 & 2.5 & 2.5 & 7 & 4 &  &     \\ \hline
$\left(r_4, lf\right)$ & & & 3.5 & 5.5 & & & 3 & 6 \\ \hline
$\left(r_4, rf\right)$ & & & 8 & 8.5 & & & 7 & 4 \\ \hline
\end{tabular}
}
\end{table}

Furthermore, the number of trains per route-train combination has to be specified to calculate the arrival rates in the Markov Chain model from Section \ref{subsec:phase_type}.
Together with the service rates, calculated from the weighted minimum headway times, queue-lengths can be estimated for every route $r \in R$.

In addition to the total number of trains $n_{\text{total}}$, the volume of traffic per combination of route-train $n_{\left(r, t\right)} \in \mathbb{R}$ depends on the share of main line traffic $p_{\text{main}} \in [0,1]$, as well as on the share of suburban trains $p_{\text{suburban}}  \in [0,1]$ and regional freight trains $p_{\text{regional freight}}  \in [0,1]$.
For the remainder of this case study, we assume a fixed rolling stock distribution of $p_{\text{suburban}} = 0.5$ for passenger and of $p_{\text{regional freight}} = 0.5$ for freight trains.

Therefore, we can formulate the functions $\theta_{\left(r, t\right)}:  \mathbb{R} \rightarrow \mathbb{R}, n_{\text{total}} \mapsto n_{r, t}$ describing the number of trains per combination for every total number of trains.

To analyse the performance capabilities of this railway junction, the introduced CTMC model for phase-type inter-arrival and service time distributions can be formulated (Section \ref{subsec:phase_type}).
Since the model's size depends on the number of phases in the service process, the variance coefficient for the service process per route $v_{S,r}$ must be calculated.

\begin{table}[!h]
\centering
\caption{Service Rates, Variance Coefficients and size of the CTMC Model (PH/PH) at different $p_{\text{main}}$ Values}
\label{tab:case_study_paras}
\resizebox{\textwidth}{!}{
\begin{tabular}{|c||c|c|c|c||c|c|c|c||c|c|}
\hline
\makecell{$p_{\text{main}}$} & \makecell{$\mu_1$} & \makecell{$\mu_2$}& \makecell{$\mu_3$} & \makecell{$\mu_4$} & \makecell{$v_{S,1}$} & \makecell{$v_{S,2}$} & \makecell{$v_{S,3}$} & \makecell{$v_{S,4}$} & \makecell{Number of \\ states} & \makecell{Number of \\ transitions} \\ \hline \hline
0.1 & 0.25 & 0.20 & 0.36 & 0.19 & 0.27 & 0.36 & 0.52 & 0.33 & \num{5.36e6} & \num{3.05e7} \\ \hline
        0.2 & 0.26 & 0.21 & 0.36 & 0.19 & 0.29 & 0.40 & 0.51 & 0.33 & \num{4.32e6} & \num{2.46e7} \\ \hline
        0.3 & 0.26 & 0.21 & 0.35 & 0.18 & 0.31 & 0.43 & 0.51 & 0.34 & \num{3.91e6} & \num{2.23e7} \\ \hline
        0.4 & 0.27 & 0.21 & 0.35 & 0.18 & 0.33 & 0.45 & 0.50 & 0.34 & \num{3.70e6} & \num{2.11e7} \\ \hline
        0.5 & 0.28 & 0.22 & 0.34 & 0.18 & 0.34 & 0.47 & 0.49 & 0.34 & \num{3.44e6} & \num{1.96e7} \\ \hline
        0.6 & 0.28 & 0.22 & 0.34 & 0.17 & 0.36 & 0.49 & 0.48 & 0.34 & \num{3.17e6} & \num{1.81e7} \\ \hline
        0.7 & 0.29 & 0.22 & 0.33 & 0.17 & 0.37 & 0.51 & 0.47 & 0.33 & \num{3.01e6} & \num{1.72e7} \\ \hline
        0.8 & 0.29 & 0.22 & 0.32 & 0.16 & 0.39 & 0.52 & 0.45 & 0.33 & \num{2.95e6} & \num{1.68e7} \\ \hline
        0.9 & 0.30 & 0.22 & 0.32 & 0.16 & 0.40 & 0.53 & 0.44 & 0.32 & \num{3.09e6} & \num{1.76e7} \\ \hline
\end{tabular}
}
\end{table}

Table \ref{tab:case_study_paras} summarizes the calculated service rates $\mu_r$, variance coefficients $v_{S,r}$ and model size for the analysed traffic distributions of $p_{\text{main}} \in \{0.1, \dots, 0.9\}$.
Additionally, the variance coefficient for the arrival process has been set to $v_A = 0.8$.

The results of those calculations are described and analysed in the following Section \ref{subsec:case_study_results}.

\subsection{Timetable Capacity Results}
\label{subsec:case_study_results}

Using the algorithm described in Section \ref{subsec:cap_deter_alg}, the performance capabilities of the railway junction with fixed $p_{\text{main}}, p_{\text{suburban}},p_{\text{regional freight}}$ traffic shares can be estimated.
For this, a lower bound of $n_{\text{total, LB}} = 1$ and an upper bound of $n_{\text{total, UB}} = 40$ has been chosen.

\begin{figure}[!htp]
    \centering
    \includegraphics[width=1\linewidth]{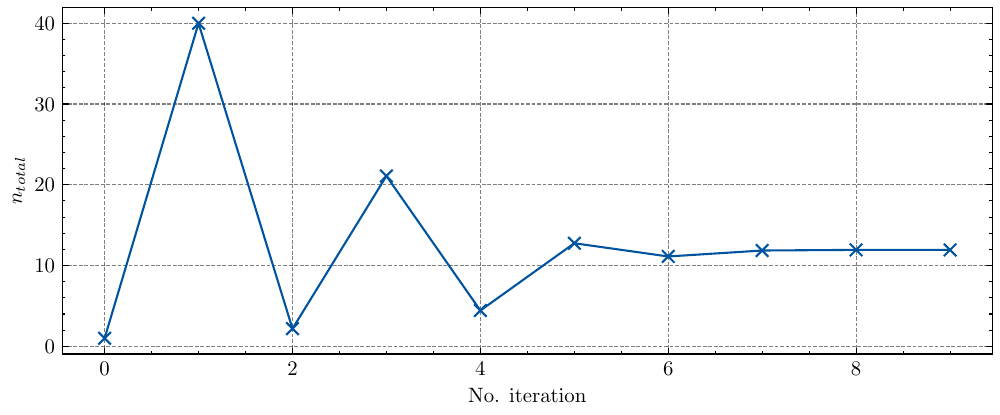}
    \caption{The number of trains $n_{\text{total}}$ in the junction with a traffic split of $p_{\text{main}}= 0.5$ for all iterations}
    \label{fig:case_study_iter}
\end{figure}

Figure \ref{fig:case_study_iter} highlights the convergence of the method for a the described example with a main line traffic share of $p_{\text{main}} = 0.5$.

The figure shows the continuous convergence of the capacity determination algorithm for a growing number of iterations.
While the given limits of $1$ and $40$ are explored in iteration 0 and 1, the algorithm terminates after $10$ calls to the model-checking, respectively after $9$ iterations at a value of $n_{\text{total}} = n_{\text{max}} = 11.93$, corresponding to the timetable capacity for the specified case. 

\begin{figure}[htp]
\centering
\begin{subfigure}{1.\textwidth}
    \centering
    \includegraphics[width=\textwidth]{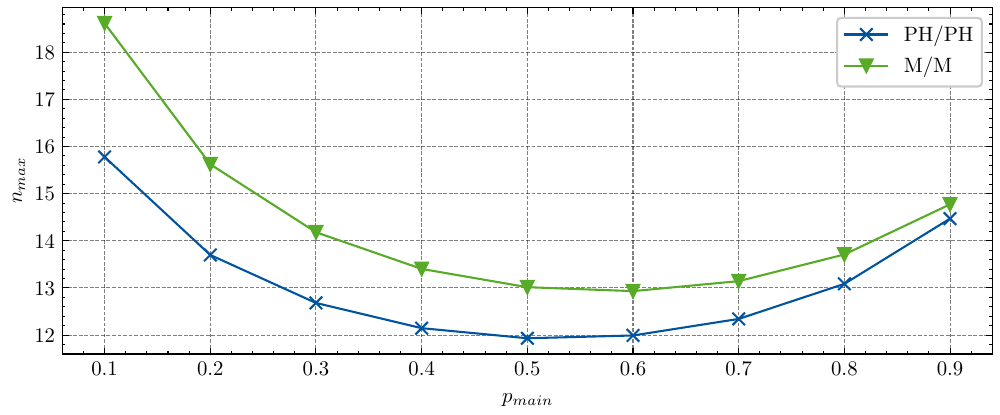}
    \caption{Timetable Capacity}
    \label{fig:case_study_results_capa}
\end{subfigure}

\hspace{0.5cm}
\begin{subfigure}{1.\textwidth}
    \centering
    \includegraphics[width=\textwidth]{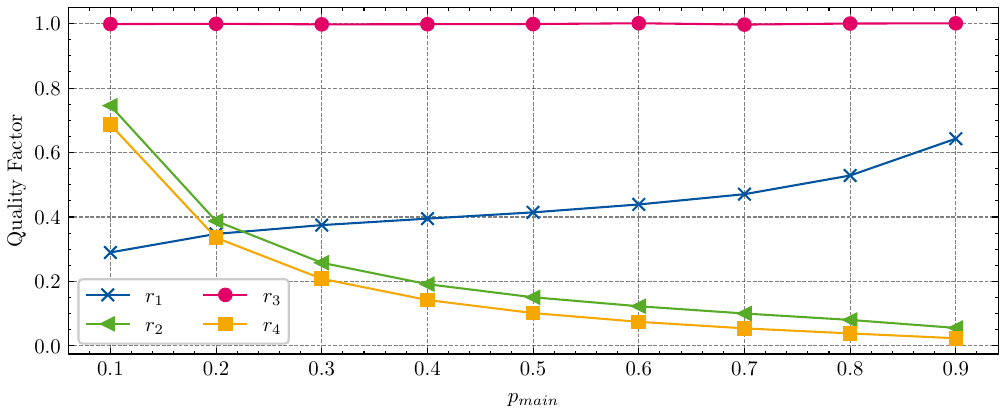}
    \caption{Quality Factor}
    \label{fig:case_study_results_qf}
\end{subfigure}
\caption{Timetable capacity $n_{\text{max}}$ of the two models PH/PH and M/M and quality factor per route for varying $p_{\text{main}}$}
\label{fig:case_study_results}
\end{figure}

The timetable capacity $n_{max}$ for other main traffic shares $p_{\text{main}} \in \{0.1, \dots, 0.9\}$ has been calculated similarly.
Figure \ref{fig:case_study_results_capa} presents the results for both the PH/PH model and the M/M model, with the latter further scaled using the Hertel approximation (see Section \ref{subsubsec:Hertel}).

The highest timetable capacities in the PH/PH model are calculated for scenarios with homogeneous traffic, with $n_{\text{max}} = 15.78$ trains per hour for $p_{\text{main}} = 0.1$ and $n_{\text{max}} = 14.47$ trains per hour for $p_{\text{main}} = 0.9$.
Conversely, the scenario with $p_{\text{main}} = 0.5$ yields the minimum timetable capacity, with $n_{\text{max}} = 11.93$ trains per hour.

Compared to the PH/PH model, the M/M model significantly overestimates  timetable capacity.
While the difference is marginal for scenarios with a high main traffic (and therefore passenger traffic) share, it increases to a full train at $p_{\text{main}} = 0.5$ and exceeds 2.5 trains at $p_{\text{main}} = 0.1$.

Furthermore, Figure \ref{fig:case_study_results_qf} depicts the quality factor for each route, which can be used to assess the infrastructure's bottleneck.
In the given example, route $r_3$ consistently has a quality factor of $1$ for every $p_{\text{main}}$, indicating that this route is the bottleneck for every described traffic scenarios.

This can be explained by the lower capacity limit for passenger traffic compared to freight traffic, and by the fact that route $r_3$ conflicts with both branching routes $r_2$ and $r_4$, which have high service times due to their exclusive utilization for freight traffic.

The influence of the traffic share on the other routes' quality is as expected: the freight traffic routes $r_2$ and $r_4$ improve their quality factors as traffic on these routes decreases, while route $r_1$ sees its quality factor increase inversely.

In summary, the analysis reveals that the introduced PH/PH model provides a more accurate estimation of timetable capacity.
Its route-based application can be used to identify the bottleneck in the infrastructure for different traffic scenarios.

\section{Discussion and Conclusion}

In this work, a novel model for performance estimation of railway junctions has been introduced, utilizing Continuous-Time Markov chains with various distribution functions for inter-arrival and service times.
The study examines the impact of using phase-type distributions compared to exponential distributions on the approximation accuracy of the models, with results compared against simulations.

Four different models, each describing distinct distributions for inter-arrival and service times, were analyzed for their approximation quality and computation time.
Special attention was given to the differences in approximation quality for various traffic distributions between the two intersecting railway lines in the analyzed railway junction.
Additionally, a case study was conducted to investigate the timetable capacity for a junction that splits a mixed traffic line into freight and passenger lines.

While the introduced model generally produces more accurate results, its computation times are significantly longer compared to models that utilize only exponential distributions.
This is due to the use of phase-type distributions, which introduce additional states into the CTMC.
As a result, the number of states becomes large even for relatively simple junctions.
Consequently, the scalability of the introduced method is highly limited.
To apply this approach to more complex railway junctions or even entire railway stations, further research is needed on decomposition techniques and alternative approximation methods.

A major parameter influencing both the model size and the timetable capacity of railway junctions is the variation coefficient of the service process.
In some chapters of this work (see Section \ref{sec:validation}), a default value was used, while in the case study, it was calculated based on minimum headway times.
For efficient use of the introduced method, it is essential to obtain an accurate estimation of this value to ensure the generation of reliable and usable results.

To assess the performance of the railway junction, queue lengths computed with CTMC models are compared to thresholds in the capacity determination algorithm (see Section \ref{subsec:cap_deter_alg}).
Although these thresholds are commonly used in practice, they were originally designed for line capacities.
Therefore, new thresholds may need to be developed to appropriately assess the quality of multi-channel systems.

The proposed method, despite its longer computation times, allows for the substitution of approximation techniques (Section \ref{sec:approx_formel}) in application scenarios that demand high accuracy. 
The introduced method demonstrates significantly shorter computation times compared to simulation approaches while producing deterministic results, making it more practical than the numerous simulation iterations needed to achieve sufficiently representative outcomes.

The proposed model, which explicitly models phase-type distributions within the Continuous-Time Markov chain, can be utilized by infrastructure managers to obtain precise, timetable-independent estimations of the capacity of railway junctions.
This enables a reduction in the reliance on computationally expensive simulations in long-term railway planning scenarios.

\section*{Acknowledgements}
This work is funded by the Deutsche Forschungsgemeinschaft (DFG, German Research Foundation) – 2236/2. Computational Experiments were performed with computing resources granted by RWTH Aachen University under projects rwth1413 and rwth1635.

\section*{CRediT authorship contribution statement}
\textbf{Tamme Emunds:} Conceptualization, Methodology, Software, Formal Analysis, Writing – original draft. \textbf{Nils Nießen:} Conceptualization, Supervision, Writing - review \& editing.

\section*{Declaration of competing interest}
The authors declare that they have no known competing financial interests of personal relationships that could have appeared to influence the work reported in this paper.


\section*{Data Availability Statement}
The processed data required to reproduce the above findings are available to download from \citet{emundsResearchDataCode2024}.

\bibliographystyle{elsarticle-harv} 
\bibliography{references}

\newpage

\appendix

\section{Capacity Determination Algorithm}
\label{app:cap_deter_algo}

In Section \ref{subsec:cap_deter_alg}, the root-finding problem has been introduced, which determines the timetable capacity of a railway junction by comparing the estimated queue-length per route $L_r$ to their respective maximum thresholds $L_{r,\text{limit}}$, while varying the total train count $n_{\text{total}}$.

Since some parameters of the railway junction, such as the arrival rates $\lambda_r$, service rates $\mu_r$ and variation coefficients $v_{A,r}, v_{S,r}$ may  depend on the total train count $n_{total}$, the CTMC model has to be formulated from inside the function call of the root-finding method.

\begin{algorithm}[!h]
\SetAlgoLined
\KwData{Railway Junction $J= \left(R,C\right)$, Minimum Headway Times $h_{(r_i, t_i), (r_j, t_j)}$,  Total Number of Trains $n_{\text{total}}$, Request Distribution $\theta$}
\KwResult{$\phi\left(n_{\text{total}}\right)$}

\ForEach{$\text{route}~ r \in R$}{%
    Determine number of trains on the route $n_r$\;
    Determine arrival rate $\lambda_r = 1/n_r$\;
    Determine share of passenger trains $p_{\text{pt}}$\;
    Determine $L_{\text{limit, r}}$\;
}

Determine frequencies of train sequences $p_{i, j}$\;

\ForEach{$\text{route}~ r \in R$}{%
    Determine service times $b_r$\;
    Determine service rates $\mu_r = 1/b_r$\;
    Determine variation coefficients $v_{S,r}$ \;
    Determine phase-type parameters $k_{A,r}, k_{S,r}, \lambda_{r,a}, \lambda_{r,b}, \mu_{r,a},  \mu_{r,a}$;
}

Build the CTMC $MC = \left(S, T\right)$\;
Get $EL_{r}\left(n_{\text{total}}\right)$\ with probabilistic model checking\;
Determine $qf_{\text{max}} \left(n_{\text{total}}\right)$\;
Determine $\phi\left(n_{\text{total}}\right)$\;
\Return{$\phi\left(n_{\text{total}}\right)$}

\caption{Inner Function for Brent's Method}
\label{alg_inner}
\end{algorithm}

Algorithm \ref{alg_inner} therefore describes the complete inner function, that computes all relevant parameters before the calculation of the quality factor and the result $\phi\left(n_{\text{total}}\right)$ for a chosen $\left(n_{\text{total}}\right)$ in an interval $\left[ n_{\text{total, LB}}, n_{\text{total, UB}}\right]$.

In the initial \texttt{foreach} loop, the number of trains per route, denoted by $n_r$, along with other computable parameters, is established. 
Subsequently, the frequencies of train sequences $p_{i, j}$ are determined, which allows for the setting of the remaining parameters in the following \texttt{foreach} loop. 
The next phase involves formulating the Continuous-Time Markov Chain (CTMC) and calculating the queue lengths $EL_{r}$. 
Finally, the difference between the maximum quality factor and the optimal ratio of $1$ is computed and returned.

With this information, Brent's method determines whether the set tolerances are met and the timetable capacity $n_{\text{max}}$ has been found. Otherwise, the interval $[a,b]$ is shortened, selecting a new bound $n_{\text{total}}$ replacing $a$ or $b$, where $\phi$ needs to be evaluated.
An example of the convergence of this method can be found in Section \ref{subsec:case_study_results}.

\newpage
\section{Performance Determination for Simulations}
\label{app:cap_deter_sim}

In Section \ref{subsec:perf_det_quality}, we conduct 100 simulations, each lasting 20 hours, for $p_{\text{main}} \in P_{\text{main}} =\{0.1, \dots, 0.9\}$ and $n_{\text{total}} \in N_{\text{total}} = \{12.00, 12.04, \dots, 19.92, 19.96\}$.

This extensive computation was necessary because the algorithm from Section \ref{subsec:cap_deter_alg} assumes monotonicity for the function $EL_{r}:  N_{\text{total}} \rightarrow \mathbb{R}, p_{\text{main}} \mapsto EL_{r} (n_{\text{total}})$.

However, since the simulation is not deterministic, even taking an average of 100 simulation runs does not guarantee a monotonic function approximation.
It is therefore possible that there exist $n_{\text{total}, 1 }, n_{\text{total}, 2 } \in N_{\text{total}}$ for a fixed $p_{\text{main}}^{\prime}  \in P_{\text{main}}$, with:
\begin{equation}
    EL_{r} (n_{\text{total}, 1 }) \geq EL_{r} (n_{\text{total}, 2})
\end{equation}
and $n_{\text{total}, 1 } < n_{\text{total}, 2}$.

Therefore, it might not be possible to find a single value $n_{\text{max}}^{\ast} \in N_{\text{total}}$ that meets the quality threshold requirement:
\begin{equation}
    EL_{r} (n_{\text{max}}^{\ast}) \leq L_{\text{limit}},
\end{equation}
where for all larger $n^{\prime} \in  N_{\text{total}}$, with $n_{\text{max}}^{\ast} < n^{\prime}$, the limit
\begin{equation}
\label{ub_sim_perf_det}
    EL_{r} (n^{\prime} ) > L_{\text{limit}}
\end{equation}
is exceeded \textbf{and} for all smaller $ n^{\prime \prime} \in  N_{\text{total}}$, with $n_{\text{max}}^{\ast} > n^{\prime \prime}$,  the limit
\begin{equation}
\label{lb_sim_perf_det}
    EL_{r} (n^{\prime \prime} ) \leq L_{\text{limit}}
\end{equation}
is not violated.

Hence, we introduce the two bounds to the capacity $n_{\text{max, LB}}^{\ast}, n_{\text{max, UB}}^{\ast} \in N_{\text{total}}$, where the lower bound $n_{\text{max, LB}}^{\ast}$ meets the condition \eqref{lb_sim_perf_det}:
For all smaller $ n^{\prime \prime} \in  N_{\text{total}}$, with $n_{\text{max, LB}}^{\ast} > n^{\prime \prime}$, the limit
\begin{equation}
    EL_{r} ( n^{\prime \prime} ) \leq L_{\text{limit}}
\end{equation}
is not violated.

Similarly, $n_{\text{max, UB}}^{\ast}$ meets the condition \eqref{ub_sim_perf_det}:
For all larger $n^{\prime} \in  N_{\text{total}}$, with $n_{\text{max, UB}}^{\ast} < n^{\prime}$, the limit
\begin{equation}
    EL_{r} (n^{\prime} ) > L_{\text{limit}}
\end{equation}
is exceeded.

These bounds are chosen to be as tight as possible, so that $n_{\text{max, LB}}^{\ast}$ represents the highest value and $n_{\text{max, UB}}^{\ast}$ represents the lowest value that satisfy the corresponding conditions.

\end{document}

%% file: rwth-colors.tex
\definecolor{rwth}   {RGB}{  0  84 159}
\definecolor{rwth-75}{RGB}{ 64 127 183}
\definecolor{rwth-50}{RGB}{142 186 229}
\definecolor{rwth-25}{RGB}{199 221 242}
\definecolor{rwth-10}{RGB}{232 241 250}

\definecolor{black}   {RGB}{  0   0   0}
\definecolor{black-75}{RGB}{100 101 103}
\definecolor{black-50}{RGB}{156 158 159}
\definecolor{black-25}{RGB}{207 209 210}
\definecolor{black-10}{RGB}{236 237 237}

\definecolor{magenta}   {RGB}{227   0 102}
\definecolor{magenta-75}{RGB}{233  96 136}
\definecolor{magenta-50}{RGB}{241 158 177}
\definecolor{magenta-25}{RGB}{249 210 218}
\definecolor{magenta-10}{RGB}{253 238 240}

\definecolor{yellow}   {RGB}{255 237   0}
\definecolor{yellow-75}{RGB}{255 240  85}
\definecolor{yellow-50}{RGB}{255 245 155}
\definecolor{yellow-25}{RGB}{255 250 209}
\definecolor{yellow-10}{RGB}{255 253 238}

\definecolor{petrol}   {RGB}{  0  97 101}
\definecolor{petrol-75}{RGB}{ 45 127 131}
\definecolor{petrol-50}{RGB}{125 164 167}
\definecolor{petrol-25}{RGB}{191 208 209}
\definecolor{petrol-10}{RGB}{230 236 236}

\definecolor{turkis}   {RGB}{  0 152 161}
\definecolor{turkis-75}{RGB}{  0 177 183}
\definecolor{turkis-50}{RGB}{137 204 207}
\definecolor{turkis-25}{RGB}{202 231 231}
\definecolor{turkis-10}{RGB}{235 246 246}

\definecolor{grun}   {RGB}{ 87 171  39}
\definecolor{grun-75}{RGB}{141 192  96}
\definecolor{grun-50}{RGB}{184 214 152}
\definecolor{grun-25}{RGB}{221 235 206}
\definecolor{grun-10}{RGB}{242 247 236}

\definecolor{maigrun}   {RGB}{189 205   0}
\definecolor{maigrun-75}{RGB}{208 217  92}
\definecolor{maigrun-50}{RGB}{224 230 154}
\definecolor{maigrun-25}{RGB}{240 243 208}
\definecolor{maigrun-10}{RGB}{249 250 237}

\definecolor{orange}   {RGB}{246 168   0}
\definecolor{orange-75}{RGB}{250 190  80}
\definecolor{orange-50}{RGB}{253 212 143}
\definecolor{orange-25}{RGB}{254 234 201}
\definecolor{orange-10}{RGB}{255 247 234}

\definecolor{rot}   {RGB}{204   7  30}
\definecolor{rot-75}{RGB}{216  92  65}
\definecolor{rot-50}{RGB}{230 150 121}
\definecolor{rot-25}{RGB}{243 205 187}
\definecolor{rot-10}{RGB}{250 235 227}

\definecolor{bordeaux}   {RGB}{161  16  53}
\definecolor{bordeaux-75}{RGB}{182  82  86}
\definecolor{bordeaux-50}{RGB}{205 139 135}
\definecolor{bordeaux-25}{RGB}{229 197 192}
\definecolor{bordeaux-10}{RGB}{245 232 229}

\definecolor{violett}   {RGB}{ 97  33  88}
\definecolor{violett-75}{RGB}{131  78 117}
\definecolor{violett-50}{RGB}{168 133 158}
\definecolor{violett-25}{RGB}{210 192 205}
\definecolor{violett-10}{RGB}{237 229 234}

\definecolor{lila}   {RGB}{122 111 172}
\definecolor{lila-75}{RGB}{155 145 193}
\definecolor{lila-50}{RGB}{188 181 215}
\definecolor{lila-25}{RGB}{222 218 235}
\definecolor{lila-10}{RGB}{242 240 247}

%% file: Media/Markov-Graphs/exponential_graph.tex
\begin{tikzpicture}[scale=2]
    \begin{scope}[every node/.style={circle,thick,draw, font=\scriptsize},
                    state/.style={minimum size=1.5cm}]
        \node [state] (s) at (0,0) {\textbf{empty}};
        \node [state] (b2) at (2,0) {\textbf{\textcolor{rwth}{s}}};
    \end{scope}
    \begin{scope}[>={Stealth[black]},
              every node/.style={fill=white,circle},
              every edge/.style={draw=black, thick, font=\scriptsize}]
        \path [->] (b2)[bend right=60] edge node {$\mu$} (s) ;
    \end{scope}
\end{tikzpicture}

%% file: Media/Markov-Graphs/phase_type_graph.tex
\begin{tikzpicture}[scale = 2]
    \begin{scope}[every node/.style={circle,thick,draw, font=\scriptsize},
                    state/.style={minimum size=1.5cm}]
        \node [state] (s) at (0,0) {\textbf{empty}};
        \node [state] (p3) at (2,0) {\textbf{\textcolor{rwth}{$p_3$}}};
        \node [state] (p2) at (4,0) {\textbf{\textcolor{rwth}{$p_2$}}};
        \node [state] (p1) at (6,0) {\textbf{\textcolor{rwth}{$p_1$}}};
        \node [state] (b2) at (8,0) {\textbf{\textcolor{rwth}{s}}};
    \end{scope}
    \begin{scope}[>={Stealth[black]},
              every node/.style={fill=white,circle},
              every edge/.style={draw=black, thick, font=\scriptsize}]
        \path [->] (b2)[bend right=60] edge node {$\mu_{p, 1}$} (p1) ;
        \path [->] (p1)[bend right=60] edge node {$\mu_{p, 2}$} (p2) ;
        \path [->] (p2)[bend right=60] edge node {$\mu_{p, 3}$} (p3) ;
        \path [->] (p3)[bend right=60] edge node {$\mu_{p, 4}$} (s) ;
        
    \end{scope}
\end{tikzpicture}

%% file: case_study_overview.tikz
\resizebox{\textwidth}{!}{
\begin{tikzpicture}
\tikzstyle{station} = [rectangle, minimum width=4.5cm, minimum height=1cm, text centered, draw=black, fill=white]
\pgfdeclarelayer{nodelayer}
\pgfdeclarelayer{edgelayer}
\pgfsetlayers{main,edgelayer, nodelayer}
	\begin{pgfonlayer}{nodelayer}
		\node [style=station] (0) at (0, 0) {Mixed Traffic Line 'A'};
		\node [style=station] (1) at (5, 0) {Considered Junction};
		\node [style=station] (2) at (13, 2) {Freight Yard 'C'};
		\node [style=station] (3) at (13, 0) {Passenger Station 'B'};
		\node [draw=none] (4) at (7, 2) {};
        \node [draw=none] (5) at (5, 2) {};
	\end{pgfonlayer}
	\begin{pgfonlayer}{edgelayer}
		\draw (0.center) to (1.center);
		\draw (1.center) -- (3.center) node [midway, fill=white, below=5pt] {main line};
		\draw (1.center) to (4.center);
		\draw (4.center) -- (2.center);
        \draw[draw=none] (5.center) -- (2.center) node [midway, fill=white, above=5pt] {branch line};
	\end{pgfonlayer}
\end{tikzpicture}
}

%% file: Phasetype_Influence.bbl
\begin{thebibliography}{55}
\expandafter\ifx\csname natexlab\endcsname\relax\def\natexlab#1{#1}\fi
\providecommand{\url}[1]{\texttt{#1}}
\providecommand{\href}[2]{#2}
\providecommand{\path}[1]{#1}
\providecommand{\DOIprefix}{doi:}
\providecommand{\ArXivprefix}{arXiv:}
\providecommand{\URLprefix}{URL: }
\providecommand{\Pubmedprefix}{pmid:}
\providecommand{\doi}[1]{\href{http://dx.doi.org/#1}{\path{#1}}}
\providecommand{\Pubmed}[1]{\href{pmid:#1}{\path{#1}}}
\providecommand{\bibinfo}[2]{#2}
\ifx\xfnm\relax \def\xfnm[#1]{\unskip,\space#1}\fi
\bibitem[{Abril et~al.(2008)Abril, Barber, Ingolotti, Salido, Tormos and Lova}]{Abril.2008}
\bibinfo{author}{Abril, M.}, \bibinfo{author}{Barber, F.}, \bibinfo{author}{Ingolotti, L.}, \bibinfo{author}{Salido, M.A.}, \bibinfo{author}{Tormos, P.}, \bibinfo{author}{Lova, A.}, \bibinfo{year}{2008}.
\newblock \bibinfo{title}{An assessment of railway capacity}.
\newblock \bibinfo{journal}{Transportation Research Part E: Logistics and Transportation Review} \bibinfo{volume}{44}, \bibinfo{pages}{774--806}.
\newblock \DOIprefix\doi{10.1016/j.tre.2007.04.001}.
\bibitem[{Asmussen et~al.(2003)Asmussen, Asmussen and Asmussen}]{asmussen2003applied}
\bibinfo{author}{Asmussen, S.}, \bibinfo{author}{Asmussen, S.}, \bibinfo{author}{Asmussen, S.}, \bibinfo{year}{2003}.
\newblock \bibinfo{title}{Applied probability and queues}. volume~\bibinfo{volume}{2}.
\newblock \bibinfo{publisher}{Springer}.
\bibitem[{Be{\v{s}}inovi{\'c} and Goverde(2018)}]{Besinovic.2018}
\bibinfo{author}{Be{\v{s}}inovi{\'c}, N.}, \bibinfo{author}{Goverde, R.M.P.}, \bibinfo{year}{2018}.
\newblock \bibinfo{title}{Capacity assessment in railway networks}, in: \bibinfo{editor}{Bornd{\"o}rfer, R.}, \bibinfo{editor}{Klug, T.}, \bibinfo{editor}{Lamorgese, L.}, \bibinfo{editor}{Mannino, C.}, \bibinfo{editor}{Reuther, M.}, \bibinfo{editor}{Schlechte, T.} (Eds.), \bibinfo{booktitle}{Handbook of Optimization in the Railway Industry}. \bibinfo{publisher}{{Springer International Publishing}}, \bibinfo{address}{Cham}. volume \bibinfo{volume}{268} of \textit{\bibinfo{series}{International Series in Operations Research {\&} Management Science}}, pp. \bibinfo{pages}{25--45}.
\newblock \DOIprefix\doi{10.1007/978-3-319-72153-8_2}.
\bibitem[{Brent(1973)}]{brent1973algorithms}
\bibinfo{author}{Brent, R.}, \bibinfo{year}{1973}.
\newblock \bibinfo{title}{Algorithms for minimization without derivatives}.
\newblock \bibinfo{journal}{Prentice-Hall, Englewood Cliffs NJ} .
\bibitem[{Burdett(2016)}]{Burdett.2016}
\bibinfo{author}{Burdett, R.L.}, \bibinfo{year}{2016}.
\newblock \bibinfo{title}{Optimisation models for expanding a railway's theoretical capacity}.
\newblock \bibinfo{journal}{European Journal of Operational Research} \bibinfo{volume}{251}, \bibinfo{pages}{783--797}.
\newblock \DOIprefix\doi{10.1016/j.ejor.2015.12.033}.
\bibitem[{Burdett and Kozan(2006)}]{Burdett.2006}
\bibinfo{author}{Burdett, R.L.}, \bibinfo{author}{Kozan, E.}, \bibinfo{year}{2006}.
\newblock \bibinfo{title}{Techniques for absolute capacity determination in railways}.
\newblock \bibinfo{journal}{Transportation Research Part B: Methodological} \bibinfo{volume}{40}, \bibinfo{pages}{616--632}.
\newblock \DOIprefix\doi{10.1016/j.trb.2005.09.004}.
\bibitem[{Cacchiani et~al.(2016)Cacchiani, Furini and Kidd}]{Cacchiani.2016}
\bibinfo{author}{Cacchiani, V.}, \bibinfo{author}{Furini, F.}, \bibinfo{author}{Kidd, M.P.}, \bibinfo{year}{2016}.
\newblock \bibinfo{title}{Approaches to a real-world train timetabling problem in a railway node}.
\newblock \bibinfo{journal}{Omega} \bibinfo{volume}{58}, \bibinfo{pages}{97--110}.
\newblock \DOIprefix\doi{10.1016/j.omega.2015.04.006}.
\bibitem[{Cacchiani and Toth(2012)}]{Cacchiani.2012}
\bibinfo{author}{Cacchiani, V.}, \bibinfo{author}{Toth, P.}, \bibinfo{year}{2012}.
\newblock \bibinfo{title}{Nominal and robust train timetabling problems}.
\newblock \bibinfo{journal}{European Journal of Operational Research} \bibinfo{volume}{219}, \bibinfo{pages}{727--737}.
\newblock \DOIprefix\doi{10.1016/j.ejor.2011.11.003}.
\bibitem[{Cox(1955)}]{cox1955use}
\bibinfo{author}{Cox, D.R.}, \bibinfo{year}{1955}.
\newblock \bibinfo{title}{A use of complex probabilities in the theory of stochastic processes}, in: \bibinfo{booktitle}{mathematical proceedings of the Cambridge philosophical society}, \bibinfo{organization}{Cambridge University Press}. pp. \bibinfo{pages}{313--319}.
\bibitem[{D'Acierno et~al.(2019)D'Acierno, Botte and Pignatiello}]{Dacierno.2019}
\bibinfo{author}{D'Acierno, L.}, \bibinfo{author}{Botte, M.}, \bibinfo{author}{Pignatiello, G.}, \bibinfo{year}{2019}.
\newblock \bibinfo{title}{A simulation-based approach for estimating railway capacity}.
\newblock \bibinfo{journal}{International Journal of Transport Development and Integration} \bibinfo{volume}{3}, \bibinfo{pages}{232--244}.
\newblock \DOIprefix\doi{10.2495/TDI-V3-N3-232-244}.
\bibitem[{David and Larry(1987)}]{david1987least}
\bibinfo{author}{David, A.}, \bibinfo{author}{Larry, S.}, \bibinfo{year}{1987}.
\newblock \bibinfo{title}{The least variable phase type distribution is erlang}.
\newblock \bibinfo{journal}{Stochastic Models} \bibinfo{volume}{3}, \bibinfo{pages}{467--473}.
\bibitem[{{DB InfraGO}(2022)}]{DBNetzAG.2009}
\bibinfo{author}{{DB InfraGO}}, \bibinfo{year}{2022}.
\newblock \bibinfo{title}{Richtlinie {Fahrwegkapazit{\"a}t}}.
\bibitem[{{DB InfraGO}(2023)}]{dbinfragoNutzungsbedingungenNetzDB2023}
\bibinfo{author}{{DB InfraGO}}, \bibinfo{year}{2023}.
\newblock \bibinfo{title}{{Nutzungsbedingungen Netz der DB Netz AG}}.
\bibitem[{De~Kort et~al.(2003)De~Kort, Heidergott and Ayhan}]{Kort.2003}
\bibinfo{author}{De~Kort, A.F.}, \bibinfo{author}{Heidergott, B.}, \bibinfo{author}{Ayhan, H.}, \bibinfo{year}{2003}.
\newblock \bibinfo{title}{A probabilistic (max, +) approach for determining railway infrastructure capacity}.
\newblock \bibinfo{journal}{European Journal of Operational Research} \bibinfo{volume}{148}, \bibinfo{pages}{644--661}.
\newblock \DOIprefix\doi{10.1016/S0377-2217(02)00467-8}.
\bibitem[{Emunds and Nie{\ss}en(2024a)}]{emundsEvaluatingRailwayJunction2024}
\bibinfo{author}{Emunds, T.}, \bibinfo{author}{Nie{\ss}en, N.}, \bibinfo{year}{2024}a.
\newblock \bibinfo{title}{Evaluating railway junction infrastructure: {{A}} queueing-based, timetable-independent analysis}.
\newblock \bibinfo{journal}{Transportation Research Part C: Emerging Technologies} \bibinfo{volume}{165}, \bibinfo{pages}{104704}.
\newblock \DOIprefix\doi{10.1016/j.trc.2024.104704}.
\bibitem[{Emunds and Nie{\ss}en(2024b)}]{emundsResearchDataCode2024}
\bibinfo{author}{Emunds, T.}, \bibinfo{author}{Nie{\ss}en, N.}, \bibinfo{year}{2024}b.
\newblock \bibinfo{title}{Research {{Data}} and {{Code}}: {{Utilizing}} phase-type distributions for queueing-based railway junction performance determination}.
\newblock \DOIprefix\doi{10.5281/ZENODO.14281662}.
\bibitem[{Fischer and Hertel(1990)}]{fischer1990bedienungsprozesse}
\bibinfo{author}{Fischer, K.}, \bibinfo{author}{Hertel, G.}, \bibinfo{year}{1990}.
\newblock \bibinfo{title}{Bedienungsprozesse im Transportwesen: Grundlagen und Anwendungen der Bedienungstheorie}.
\newblock \bibinfo{publisher}{Transpress-Verlag}.
\bibitem[{Goverde(2007)}]{Goverde.2007}
\bibinfo{author}{Goverde, R.M.}, \bibinfo{year}{2007}.
\newblock \bibinfo{title}{Railway timetable stability analysis using max-plus system theory}.
\newblock \bibinfo{journal}{Transportation Research Part B: Methodological} \bibinfo{volume}{41}, \bibinfo{pages}{179--201}.
\newblock \DOIprefix\doi{10.1016/j.trb.2006.02.003}.
\bibitem[{Goverde et~al.(2013)Goverde, Corman and D'Ariano}]{Goverde.2013}
\bibinfo{author}{Goverde, R.M.}, \bibinfo{author}{Corman, F.}, \bibinfo{author}{D'Ariano, A.}, \bibinfo{year}{2013}.
\newblock \bibinfo{title}{Railway line capacity consumption of different railway signalling systems under scheduled and disturbed conditions}.
\newblock \bibinfo{journal}{Journal of Rail Transport Planning {\&} Management} \bibinfo{volume}{3}, \bibinfo{pages}{78--94}.
\newblock \DOIprefix\doi{10.1016/j.jrtpm.2013.12.001}.
\bibitem[{Gudehus(1976)}]{gudehusStaueffekteVorTransportknoten1976a}
\bibinfo{author}{Gudehus, T.}, \bibinfo{year}{1976}.
\newblock \bibinfo{title}{{Staueffekte vor Transportknoten}}.
\newblock \bibinfo{journal}{Zeitschrift f{\"u}r Operations Research} \bibinfo{volume}{20}, \bibinfo{pages}{B207--B252}.
\newblock \DOIprefix\doi{10.1007/BF01918395}.
\bibitem[{Hansen and Pachl(2014)}]{hansen2014railway}
\bibinfo{author}{Hansen, I.}, \bibinfo{author}{Pachl, J.}, \bibinfo{year}{2014}.
\newblock \bibinfo{title}{Railway timetabling and operations: Analysis, modelling, optimisation, simulation, performance, evaluation}.
\newblock \bibinfo{publisher}{Eurail press}.
\bibitem[{Harrod(2009)}]{Harrod.2009}
\bibinfo{author}{Harrod, S.}, \bibinfo{year}{2009}.
\newblock \bibinfo{title}{Capacity factors of a mixed speed railway network}.
\newblock \bibinfo{journal}{Transportation Research Part E: Logistics and Transportation Review} \bibinfo{volume}{45}, \bibinfo{pages}{830--841}.
\newblock \DOIprefix\doi{10.1016/j.tre.2009.03.004}.
\bibitem[{Hensel et~al.(2022)Hensel, Junges, Katoen, Quatmann and Volk}]{Hensel.2022}
\bibinfo{author}{Hensel, C.}, \bibinfo{author}{Junges, S.}, \bibinfo{author}{Katoen, J.P.}, \bibinfo{author}{Quatmann, T.}, \bibinfo{author}{Volk, M.}, \bibinfo{year}{2022}.
\newblock \bibinfo{title}{The probabilistic model checker storm}.
\newblock \bibinfo{journal}{International Journal on Software Tools for Technology Transfer} \bibinfo{volume}{24}, \bibinfo{pages}{589--610}.
\newblock \DOIprefix\doi{10.1007/s10009-021-00633-z}.
\bibitem[{Hertel(1984)}]{hertel1984exakte}
\bibinfo{author}{Hertel, G.}, \bibinfo{year}{1984}.
\newblock \bibinfo{title}{Exakte l{\"o}sung zur berechnung der wartegleiszahl vor im einrichtungsbetrieb befahrenen streckengleisen bei nicht-poisson-ank{\"u}nften (g/m/1-wartesystem)}.
\newblock \bibinfo{journal}{Wissenschaftliche Zeitschrift der Hochschule f{\"u}r Verkehrswesen} \bibinfo{volume}{31}, \bibinfo{pages}{195--205}.
\bibitem[{Jensen et~al.(2020)Jensen, Schmidt and Nielsen}]{Jensen.2020}
\bibinfo{author}{Jensen, L.W.}, \bibinfo{author}{Schmidt, M.}, \bibinfo{author}{Nielsen, O.A.}, \bibinfo{year}{2020}.
\newblock \bibinfo{title}{Determination of infrastructure capacity in railway networks without the need for a fixed timetable}.
\newblock \bibinfo{journal}{Transportation Research Part C: Emerging Technologies} \bibinfo{volume}{119}, \bibinfo{pages}{102751}.
\newblock \DOIprefix\doi{10.1016/j.trc.2020.102751}.
\bibitem[{Junges and Volk(2023)}]{SebastianJunges.2023}
\bibinfo{author}{Junges, S.}, \bibinfo{author}{Volk, M.}, \bibinfo{year}{2023}.
\newblock \bibinfo{title}{Stormpy - python bindings for storm: Version 1.7.0}.
\newblock \URLprefix \url{https://moves-rwth.github.io/stormpy/index.html}.
\bibitem[{Kendall(1953)}]{Kendall.1953}
\bibinfo{author}{Kendall, D.G.}, \bibinfo{year}{1953}.
\newblock \bibinfo{title}{Stochastic processes occurring in the theory of queues and their analysis by the method of the imbedded markov chain}.
\newblock \bibinfo{journal}{The Annals of Mathematical Statistics} , \bibinfo{pages}{338--354}.
\bibitem[{Kingman(1961)}]{kingmanSingleServerQueue1961}
\bibinfo{author}{Kingman, J.F.C.}, \bibinfo{year}{1961}.
\newblock \bibinfo{title}{The single server queue in heavy traffic}.
\newblock \bibinfo{journal}{Mathematical Proceedings of the Cambridge Philosophical Society} \bibinfo{volume}{57}, \bibinfo{pages}{902--904}.
\newblock \DOIprefix\doi{10.1017/S0305004100036094}.
\bibitem[{Kwiatkowska et~al.(2011)Kwiatkowska, Norman and Parker}]{KNP11}
\bibinfo{author}{Kwiatkowska, M.}, \bibinfo{author}{Norman, G.}, \bibinfo{author}{Parker, D.}, \bibinfo{year}{2011}.
\newblock \bibinfo{title}{{PRISM} 4.0: Verification of probabilistic real-time systems}, in: \bibinfo{editor}{Gopalakrishnan, G.}, \bibinfo{editor}{Qadeer, S.} (Eds.), \bibinfo{booktitle}{Proc. 23rd International Conference on Computer Aided Verification (CAV'11)}, \bibinfo{publisher}{Springer}. pp. \bibinfo{pages}{585--591}.
\bibitem[{Liao et~al.(2021)Liao, Li, Miao and Corman}]{Liao.2021}
\bibinfo{author}{Liao, Z.}, \bibinfo{author}{Li, H.}, \bibinfo{author}{Miao, J.}, \bibinfo{author}{Corman, F.}, \bibinfo{year}{2021}.
\newblock \bibinfo{title}{Railway capacity estimation considering vehicle circulation: Integrated timetable and vehicles scheduling on hybrid time-space networks}.
\newblock \bibinfo{journal}{Transportation Research Part C: Emerging Technologies} \bibinfo{volume}{124}, \bibinfo{pages}{102961}.
\newblock \DOIprefix\doi{10.1016/j.trc.2020.102961}.
\bibitem[{Lusby et~al.(2011)Lusby, Larsen, Ryan and Ehrgott}]{Lusby.2011}
\bibinfo{author}{Lusby, R.}, \bibinfo{author}{Larsen, J.}, \bibinfo{author}{Ryan, D.}, \bibinfo{author}{Ehrgott, M.}, \bibinfo{year}{2011}.
\newblock \bibinfo{title}{Routing trains through railway junctions: A new set-packing approach}.
\newblock \bibinfo{journal}{Transportation Science} \bibinfo{volume}{45}, \bibinfo{pages}{228--245}.
\newblock \DOIprefix\doi{10.1287/trsc.1100.0362}.
\bibitem[{Mussone and {Wolfler Calvo}(2013)}]{Mussone.2013}
\bibinfo{author}{Mussone, L.}, \bibinfo{author}{{Wolfler Calvo}, R.}, \bibinfo{year}{2013}.
\newblock \bibinfo{title}{An analytical approach to calculate the capacity of a railway system}.
\newblock \bibinfo{journal}{European Journal of Operational Research} \bibinfo{volume}{228}, \bibinfo{pages}{11--23}.
\newblock \DOIprefix\doi{10.1016/j.ejor.2012.12.027}.
\bibitem[{Nie{\ss}en(2008)}]{Niessen.2008}
\bibinfo{author}{Nie{\ss}en, N.}, \bibinfo{year}{2008}.
\newblock \bibinfo{title}{Leistungskenngr{\"o}{\ss}en f{\"u}r Gesamtfahrstra{\ss}enknoten}.
\newblock Ph.D. thesis. {Verkehrswissenschaftliches Institut der RWTH Aachen} (german only).
\bibitem[{{Nie{\ss}en}(2013)}]{Niessen.2013}
\bibinfo{author}{{Nie{\ss}en}, N.}, \bibinfo{year}{2013}.
\newblock \bibinfo{title}{Waiting and loss probabilities for route nodes}, in: \bibinfo{booktitle}{International Conference on Railway Operations Modelling and Analysis (RailCopenhagen)}.
\bibitem[{Palmer et~al.(2019)Palmer, Knight, Harper and Hawa}]{ciwarticle}
\bibinfo{author}{Palmer, G.I.}, \bibinfo{author}{Knight, V.A.}, \bibinfo{author}{Harper, P.R.}, \bibinfo{author}{Hawa, A.L.}, \bibinfo{year}{2019}.
\newblock \bibinfo{title}{Ciw: An open-source discrete event simulation library}.
\newblock \bibinfo{journal}{Journal of Simulation} \bibinfo{volume}{13}, \bibinfo{pages}{68--82}.
\newblock \DOIprefix\doi{10.1080/17477778.2018.1473909}.
\bibitem[{Parker et~al.(2000)Parker, Norman and Kwiatkowska}]{DaveParkerGethinNormanMartaKwiatkowska.2000}
\bibinfo{author}{Parker, D.}, \bibinfo{author}{Norman, G.}, \bibinfo{author}{Kwiatkowska, M.}, \bibinfo{year}{2000}.
\newblock \bibinfo{title}{Prism}.
\newblock \URLprefix \url{https://www.prismmodelchecker.org/}.
\bibitem[{Potthoff(1970)}]{GerhartPotthoff.1970}
\bibinfo{author}{Potthoff, G.}, \bibinfo{year}{1970}.
\newblock \bibinfo{title}{Verkehrsstr{\"o}mungslehre - Die Zugfolge auf Strecken und in Bahnh{\"o}fen}.
\newblock \bibinfo{publisher}{Transpress (german only)}.
\bibitem[{{Python Software Foundation}(2022)}]{PythonSoftwareFoundation.2022}
\bibinfo{author}{{Python Software Foundation}}, \bibinfo{year}{2022}.
\newblock \bibinfo{title}{Python 3.10.9}.
\newblock \URLprefix \url{https://docs.python.org/release/3.10.9/index.html}.
\bibitem[{Schmitz et~al.(2017)Schmitz, Weik, Zieger, Nie{\ss}en and Schmeink}]{Schmitz.2017}
\bibinfo{author}{Schmitz, C.}, \bibinfo{author}{Weik, N.}, \bibinfo{author}{Zieger, S.}, \bibinfo{author}{Nie{\ss}en, N.}, \bibinfo{author}{Schmeink, A.}, \bibinfo{year}{2017}.
\newblock \bibinfo{title}{Markov models for the performance analysis of railway networks}, in: \bibinfo{booktitle}{International Conference on Railway Operations Modelling and Analysis (RailLille)}, \bibinfo{address}{Lille, France}. p.~\bibinfo{pages}{23}.
\bibitem[{Schwanh{\"a}u{\ss}er(1974)}]{Schwanhauer.1974}
\bibinfo{author}{Schwanh{\"a}u{\ss}er, W.}, \bibinfo{year}{1974}.
\newblock \bibinfo{title}{Die Bemessung der Pufferzeiten im Fahrplangef{\"u}ge der Eisenbahn}.
\newblock Ph.D. thesis. {Verkehrswissenschaftliches Institut der Rheinisch-Westf{\"a}lischen Technischen Hochschule Aachen} (german only).
\bibitem[{Schwanh{\"a}u{\ss}er(1978)}]{Schwanhauer.1978}
\bibinfo{author}{Schwanh{\"a}u{\ss}er, W.}, \bibinfo{year}{1978}.
\newblock \bibinfo{title}{Die ermittlung der leistungsf{\"a}higkeit von gro{\ss}en fahrstra{\ss}enknoten und von teilen des eisenbahnetzes}.
\newblock \bibinfo{journal}{Archiv f{\"u}r Eisenbahntechnik} \bibinfo{volume}{1978}, \bibinfo{pages}{7--18}.
\bibitem[{{Schwanhäußer} and Schultze(1982)}]{Schwanhauer.1982}
\bibinfo{author}{{Schwanhäußer}, W.}, \bibinfo{author}{Schultze, K.}, \bibinfo{year}{1982}.
\newblock \bibinfo{title}{Ermittlung von Qualit{\"a}tsma{\ss}st{\"a}ben f{\"u}r die Berechnung der Leistungsf{\"a}higkeit eines Streckenabschnittes und Entwicklung eines Rechenverfahrens zur Ermittlung von Endversp{\"a}tungen: Forschungsarbeit f{\"u}r die Deutsche Bundesbahn}.
\newblock \bibinfo{publisher}{(work not formally published, german only)}.
\bibitem[{Sommereder(2011)}]{sommereder2011modelling}
\bibinfo{author}{Sommereder, M.}, \bibinfo{year}{2011}.
\newblock \bibinfo{title}{Modelling of Queueing Systems with Markov Chains: An Introduction to Basic and Advanced Modelling Techniques}.
\newblock \bibinfo{publisher}{BoD--Books on Demand}.
\bibitem[{{Team SimPy Revision}(2023)}]{TeamSimPyRevision.2023}
\bibinfo{author}{{Team SimPy Revision}}, \bibinfo{year}{2023}.
\newblock \bibinfo{title}{Simpy 4.0.1}.
\newblock \URLprefix \url{https://simpy.readthedocs.io/en/4.0.1/}.
\bibitem[{{{The Ciw library developers}}(2024)}]{ciwpython}
\bibinfo{author}{{{The Ciw library developers}}}, \bibinfo{year}{2024}.
\newblock \bibinfo{title}{Ciw: v3.2.2}.
\newblock \DOIprefix\doi{10.5281/zenodo.12667778}.
\bibitem[{UIC(2004)}]{UIC.2004}
\bibinfo{author}{UIC}, \bibinfo{year}{2004}.
\newblock \bibinfo{title}{Code 406 - \uppercase{C}apacity}.
\bibitem[{UIC(2013)}]{uicCode406Capacity2013}
\bibinfo{author}{UIC}, \bibinfo{year}{2013}.
\newblock \bibinfo{title}{Code 406 - {{Capacity}}}.
\bibitem[{Van~Rossum and Drake~Jr(1995)}]{van1995python}
\bibinfo{author}{Van~Rossum, G.}, \bibinfo{author}{Drake~Jr, F.L.}, \bibinfo{year}{1995}.
\newblock \bibinfo{title}{Python tutorial}.
\newblock \bibinfo{publisher}{Centrum voor Wiskunde en Informatica Amsterdam, The Netherlands}.
\bibitem[{Virtanen et~al.(2020)Virtanen, Gommers, Oliphant, Haberland, Reddy, Cournapeau, Burovski, Peterson, Weckesser, Bright, {van der Walt}, Brett, Wilson, Millman, Mayorov, Nelson, Jones, Kern, Larson, Carey, Polat, Feng, Moore, {VanderPlas}, Laxalde, Perktold, Cimrman, Henriksen, Quintero, Harris, Archibald, Ribeiro, Pedregosa, {van Mulbregt} and {SciPy 1.0 Contributors}}]{2020SciPy-NMeth}
\bibinfo{author}{Virtanen, P.}, \bibinfo{author}{Gommers, R.}, \bibinfo{author}{Oliphant, T.E.}, \bibinfo{author}{Haberland, M.}, \bibinfo{author}{Reddy, T.}, \bibinfo{author}{Cournapeau, D.}, \bibinfo{author}{Burovski, E.}, \bibinfo{author}{Peterson, P.}, \bibinfo{author}{Weckesser, W.}, \bibinfo{author}{Bright, J.}, \bibinfo{author}{{van der Walt}, S.J.}, \bibinfo{author}{Brett, M.}, \bibinfo{author}{Wilson, J.}, \bibinfo{author}{Millman, K.J.}, \bibinfo{author}{Mayorov, N.}, \bibinfo{author}{Nelson, A.R.J.}, \bibinfo{author}{Jones, E.}, \bibinfo{author}{Kern, R.}, \bibinfo{author}{Larson, E.}, \bibinfo{author}{Carey, C.J.}, \bibinfo{author}{Polat, {\.I}.}, \bibinfo{author}{Feng, Y.}, \bibinfo{author}{Moore, E.W.}, \bibinfo{author}{{VanderPlas}, J.}, \bibinfo{author}{Laxalde, D.}, \bibinfo{author}{Perktold, J.}, \bibinfo{author}{Cimrman, R.}, \bibinfo{author}{Henriksen, I.}, \bibinfo{author}{Quintero, E.A.}, \bibinfo{author}{Harris, C.R.}, \bibinfo{author}{Archibald, A.M.}, \bibinfo{author}{Ribeiro, A.H.},
  \bibinfo{author}{Pedregosa, F.}, \bibinfo{author}{{van Mulbregt}, P.}, \bibinfo{author}{{SciPy 1.0 Contributors}}, \bibinfo{year}{2020}.
\newblock \bibinfo{title}{{{SciPy} 1.0: Fundamental Algorithms for Scientific Computing in Python}}.
\newblock \bibinfo{journal}{Nature Methods} \bibinfo{volume}{17}, \bibinfo{pages}{261--272}.
\newblock \DOIprefix\doi{10.1038/s41592-019-0686-2}.
\bibitem[{Weik(2020)}]{Weik.2020PhD}
\bibinfo{author}{Weik, N.}, \bibinfo{year}{2020}.
\newblock \bibinfo{title}{Long-Term Capacity Planning of Railway Infrastructure – A Stochastic Approach Capturing Infrastructure Unavailability}.
\newblock Ph.D. thesis. {RWTH Aachen University}.
\newblock \DOIprefix\doi{10.18154/RWTH-2020-06771}.
\bibitem[{Weik and Nie{\ss}en(2017)}]{Weik.2017}
\bibinfo{author}{Weik, N.}, \bibinfo{author}{Nie{\ss}en, N.}, \bibinfo{year}{2017}.
\newblock \bibinfo{title}{A quasi-birth-and-death process approach for integrated capacity and reliability modeling of railway systems}.
\newblock \bibinfo{journal}{Journal of Rail Transport Planning {\&} Management} \bibinfo{volume}{7}, \bibinfo{pages}{114--126}.
\newblock \DOIprefix\doi{10.1016/j.jrtpm.2017.06.001}.
\bibitem[{Wendler(2007)}]{Wendler.2007}
\bibinfo{author}{Wendler, E.}, \bibinfo{year}{2007}.
\newblock \bibinfo{title}{The scheduled waiting time on railway lines}.
\newblock \bibinfo{journal}{Transportation Research Part B: Methodological} \bibinfo{volume}{41}, \bibinfo{pages}{148--158}.
\newblock \DOIprefix\doi{10.1016/j.trb.2006.02.009}.
\bibitem[{Yaghini et~al.(2014)Yaghini, Nikoo and Ahadi}]{Yaghini.2014}
\bibinfo{author}{Yaghini, M.}, \bibinfo{author}{Nikoo, N.}, \bibinfo{author}{Ahadi, H.R.}, \bibinfo{year}{2014}.
\newblock \bibinfo{title}{An integer programming model for analysing impacts of different train types on railway line capacity}.
\newblock \bibinfo{journal}{Transport} \bibinfo{volume}{29}, \bibinfo{pages}{28--35}.
\newblock \DOIprefix\doi{10.3846/16484142.2014.894938}.
\bibitem[{Zieger et~al.(2018)Zieger, Weik and Nie{\ss}en}]{Zieger.2018}
\bibinfo{author}{Zieger, S.}, \bibinfo{author}{Weik, N.}, \bibinfo{author}{Nie{\ss}en, N.}, \bibinfo{year}{2018}.
\newblock \bibinfo{title}{The influence of buffer time distributions in delay propagation modelling of railway networks}.
\newblock \bibinfo{journal}{Journal of Rail Transport Planning {\&} Management} \bibinfo{volume}{8}, \bibinfo{pages}{220--232}.
\newblock \DOIprefix\doi{10.1016/j.jrtpm.2018.09.001}.
\bibitem[{Zwaneveld et~al.(1996)Zwaneveld, Kroon, Romeijn, Salomon, Dauz{\`e}re-P{\'e}r{\`e}s, {van Hoesel} and Ambergen}]{Zwaneveld.1996}
\bibinfo{author}{Zwaneveld, P.J.}, \bibinfo{author}{Kroon, L.G.}, \bibinfo{author}{Romeijn, H.E.}, \bibinfo{author}{Salomon, M.}, \bibinfo{author}{Dauz{\`e}re-P{\'e}r{\`e}s, S.}, \bibinfo{author}{{van Hoesel}, S.P.M.}, \bibinfo{author}{Ambergen, H.W.}, \bibinfo{year}{1996}.
\newblock \bibinfo{title}{Routing trains through railway stations: Model formulation and algorithms}.
\newblock \bibinfo{journal}{Transportation Science} \bibinfo{volume}{30}, \bibinfo{pages}{181--194}.
\newblock \DOIprefix\doi{10.1287/trsc.30.3.181}.

\end{thebibliography}
